\documentclass[12pt]{article}

\usepackage{layout}
\usepackage[textwidth=460pt, textheight=630pt,voffset=15pt]{geometry}
\usepackage[T1]{fontenc}

\usepackage{graphicx}%
\usepackage{multirow}%
\usepackage{amsmath,amssymb,amsfonts}%
\usepackage{amsthm}%
\usepackage{mathrsfs}%
\usepackage{mathtools}
\usepackage{stmaryrd} % for \llbrackets
\usepackage{accents} % for \llbrackets
\usepackage{tensor}
\usepackage{enumitem}

\usepackage[title]{appendix}%
\usepackage{xcolor}%
\usepackage{textcomp}%
\usepackage{manyfoot}%
\usepackage{booktabs}%
\usepackage{algorithm}%
\usepackage{algorithmicx}%
\usepackage{algpseudocode}%x
\usepackage{listings}%
\usepackage{subcaption}
\usepackage{etoolbox}
\usepackage{multicol}

\usepackage[pdftex,colorlinks=true, pdfstartview=FitV, linkcolor= Mlink, citecolor=Mlink, urlcolor= Mlink, hyperindex=true, hyperfigures=true]{hyperref}
\usepackage{authblk}
 \usepackage[numbers,sort&compress]{natbib}
\input{aastexv6.sty}% For ADS abbreviations
\input{Short_cut_2.0.sty}

\title{\bf Topologically modified Einstein equation: \\
a solution with singularities on $\mS^3$}

\author[1,2]{{Quentin} {Vigneron}\footnote{\href{mailto:quentin.vigneron@umk.pl}{quentin.vigneron@umk.pl}}}
\author[1]{{\'Aron} {Szab\'o}\footnote{\href{mailto:aron.szabo@v.umk.pl}{aron.szabo@v.umk.pl}}}
\author[3]{{Pierre} {Mourier}\footnote{\href{mailto:pierre.mourier@uib.es}{pierre.mourier@uib.es}}}

\affil[1]{\small\it{Institute of Astronomy, Faculty of Physics, Astronomy and Informatics}, {Nicolaus Copernicus University}, {{Grudzi{\k{a}}dzka 5}, {Toru\'n}, {87-100} {Poland}}}
\affil[2]{\small\it{ENS de Lyon, CRAL UMR5574, Universit\'e Claude Bernard Lyon 1, CNRS, Lyon, F-69007, France}}
\affil[3]{\small\it{Departament de F\'isica, Universitat de les Illes Balears, IAC3 – IEEC, Crta. Valldemossa km 7.5, E-07122 Palma, Spain}}
\date{\vspace{-.5cm}(\today)}

\begin{document}
%\layout

\maketitle

\begin{abstract}{
\noindent {Vigneron [{\it Foundations of Physics}, {\bf 54}, 15, (\href{https://arxiv.org/abs/2204.13980}{2024})] recently} proposed a modification of general relativity in which a non-dynamical term related {to the spatial topology} is introduced in the Einstein equation. 
The original motivation for this theory is to allow for the non-relativistic limit to exist in any physical topology. In the present paper, we derive {a first inhomogeneous exact vacuum solution of this theory for a spherical topology}, assuming staticity and spherical symmetry. The metric represents a black hole %(with positive Komar mass) 
 and a {repulsive singularity} %(with negative and opposite Komar mass) 
at opposite poles of a {3-sphere}. The solution is similar to the Schwarzschild metric, but the spacelike infinity is cut, and replaced by a {repulsive singularity} at finite distance, implying that the spacelike hypersurfaces {have finite volume}, and the total mass is zero. 
{We discuss how this solution paves the way to massive, non-static solutions of this theory, more directly relevant for cosmology.}

}
\end{abstract}

%\keywords{keyword1, Keyword2, Keyword3, Keyword4}



\newpage
\section{Introduction and motivations}

In a recent paper \citep{2024_Vigneron}, we showed why general relativity (GR) is not compatible with Newtonian gravitation (understood as Galilean invariant gravitation, see \cite{1976_Kunzle, 2021_Vigneron}) on a 3-sphere, and more generally on any closed non-Euclidean topology, i.e. for which the spatial Riemannian covering space is not $\mE^3$ (e.g. $\mS^3$, $\mH^3$, $\mathbb R\times\mS^2$, ... \citep{1995_La_Lu}). In that same paper, we gave arguments to interpret this result as a ``needle in the foot'' of general relativity when non-Euclidean topologies are considered, and we minimally modified the Einstein equation such that it becomes compatible with Newtonian gravitation in any spatial topology. In that modification, a non-dynamical term related to the topology of the Universe is added in the Einstein equation. This term depends on a reference non-dynamical connection, additional to the physical metric connection. 
%This is a large scales modifications of the theory that keeps the fundamental principles of general relativity like the equivalence principle. 
As will be shown in Section~\ref{sec:bi-connection}, this modification is conceptually close to GR, e.g. the equivalence principle is still present, but differs at larges scales where the effects of topology become important. The two theories are equivalent for Euclidean topologies, and differ for any other type of topology.
%As will be presented in Section~\ref{sec:bi-connection}, this theory can be seen as a subcase of metric affine gravity (e.g. \citep[][]{1995_Hehl_et_al}) where the non-metric connection is fixed, symmetric, (non-flat) and depends on the spatial topology. 
Different names could be attributed to this modification: e.g. ``Topological Symmetric Metric Affine Gravity'' as it can be seen as a subcase of metric affine gravity (e.g. \citep[][]{1995_Hehl_et_al}); ``Topologically Corrected General Relativity'' as it aims at correcting GR for its lack of non-relativistic limit depending on topology. In the present paper, we choose the second proposition and abbreviate it as ``topo-GR'', which highlights the fact that the theory remains close to GR.\saut%, and is even equivalent in the case of Euclidean topologies.\saut

In \citep{2023_Vigneron_et_al_b}, we derived the homogeneous and isotropic solution of this theory, along with its first order perturbations with the aim of testing the theory with cosmological data. The goal of the present paper is to study an exact inhomogeneous solution in a non-Euclidean topology: a vacuum spherically symmetric metric on a $\mathbb R\times\mS^3$ spacetime topology. %Such a metric is especially interesting as it would represent black holes in a closed Universe
Such a solution does not exist in general relativity due to Birkhoff's theorem \citep{2011_Uzan_et_al}.
In this paper, we will focus on a static solution. As will be shown, it represents an attractive singularity shielded by a horizon (black hole) and a repulsive naked singularity placed at opposite poles of a 3-sphere. The presence of the repulsive naked singularity restricts the physical relevance of this exact solution. However, as argued in Section~\ref{sec:BH_Alone}, removing {this singularity}, i.e. having only one black hole on $\mS^3$, requires dropping the staticity hypothesis. This is beyond the scope of this paper, but constitutes one of the main prospect for a generalization of the solution.
%is its main prospect. 
Still, this solution allows us to further understand the differences brought by {\MyTheory} compared to GR when non-trivial spatial topologies are considered, as discussed in Section~\ref{sec:Discussions}.\saut

The outline of the paper is the following: in Section~\ref{sec:bi-connection}, we describe the {\MyTheory} theory introduced in \citep{2024_Vigneron}; in Section~\ref{sec:The_solution}, we provide the static vacuum spherically symmetric solution of the theory and analyse its main properties (Penrose--Carter diagram, stable orbits, mass, topology); in Section~\ref{sec:Discussions}, we put in perspective this solution with the description of singularities on $\mS^3$ in GR, and we discuss the main prospects of the paper for the study of {\MyTheory}. We conclude in Section~\ref{sec:Conclusion}.

\newpage
\section{Topologically Corrected General Relativity}
\label{sec:bi-connection}

We summarise in this section the theory developed in \citep{2024_Vigneron}, which we name ``Topogically Corrected General Relativity'' (topo-GR) in the present paper. {We present here only what is needed for this paper. The reader
may find a more exhaustive description of the theory in \citep{2024_Vigneron}.} The first section~\ref{sec_Topo} presents the mathematical context related to topology, while Sections~\ref{sec_Rbar} and~\ref{sec_Field} present the fields involved in the theory and the physical equations.

\subsection{Classification of 3-dimensional closed topologies}
\label{sec_Topo}

The classification of closed 2-dimensional surfaces states that any closed connected surface is a connected sum of prime surfaces whose Riemannian universal covering space (or surface) is either the Euclidean plane $\mE^2$, the 2-sphere $\mS^2$ or the hyperbolic plane $\mH^2$. In three dimensions, a similar classification exists for closed {connected} spaces, called the Thurston classification. But instead of featuring three classes, this classification has eight distinct classes with Riemannian universal covering spaces being $\tilde\Sigma  \in \{\mE^3, \ \mS^3,\ \mH^3, \ \mR\times\mS^2, \ \mR\times\mH^2, \ Nil, \ Sol, \ \tilde{SL}(2,\mR)\}$ (see \citep{1995_La_Lu} for an overview of topology in the context of cosmology). As for the 2-dimensional case, a 3-manifold $\Sigma$ whose Riemannian covering space $\tilde\Sigma$ is $\mE^3$, $\mS^3$ or $\mH^3$ is said to have, respectively, a Euclidean, spherical or hyperbolic topology.\saut

Note that the right mathematical vocabulary that should be used is ``geometry'' instead of ``topology''. The reason is that the Thurston classification is related to the classification of (geometric) G-structures on compact 3-manifolds, which implies a classification of 3-dimensional closed topological spaces. However, ``geometry'' as usually considered in general relativity or cosmology has a different meaning. Especially, ``Euclidean geometry'' usually refers to having a flat physical metric. This is not the case here. So, we prefer to use ``topology'' instead of ``geometry'', highlighting the fact that ``Euclidean'', ``Spherical'' or ``Hyperbolic'' refer first to topological properties rather than curvature properties.\saut

Finally, in the context of the theory of \citep{2024_Vigneron}, globally hyperbolic spacetimes 4-manifolds $\CM$ with closed and connected spatial leaves are always considered, i.e. $\CM = \mathbb{R}\times\Sigma$ with $\Sigma$ a closed 3-manifold. Therefore, the choice of spacetime covering space $\tilde \CM$ directly corresponds to a choice of spatial covering space $\tilde\Sigma$ because $\tilde\CM = \mR\times\tilde\Sigma$ (see Theorem A (ii) in \citep[][]{2018_Hausmann_et_al}). Consequently, the spatial hypersurfaces are described by the Thurston classification presented above.

\vspace{-.2cm}
\subsection{The reference curvature}
\label{sec_Rbar}

The theory constructed in \citep{2024_Vigneron} is a bi-connection theory: it is defined on a 4-manifold $\CM$ equipped with a (physical) Lorentzian metric $g_{\alpha\beta}$ with its Levi-Civita connection $\T\nabla$ and defining the physical spacetime Riemann curvature ${R}^{\alpha}{}_{\beta\mu\nu}$; and a non-dynamical, symmetric (reference) connection $\T{\bar \nabla}$, defining a reference spacetime Riemann curvature $\bar{R}^{\alpha}{}_{\beta\mu\nu}$. This second connection is non-dynamical in the sense that it is the same for any metric and energy-momentum tensor, for a given topology. {The reference connection $\T{\bar\nabla}$ is not necessarily related to a metric.}\saut

The key feature of {\MyTheory} is that $\bar{\T\nabla}$ is chosen to depend on the spacetime topology. More precisely, it depends on the spacetime covering space, and therefore on the spatial covering space due to the closed connected global hyperbolicity hypothesis made in the previous section. The reference curvature is chosen to be the curvature of the (external Whitney) sum of any connection on $\mR$ and a connection on $\Sigma$, which comes from the universal cover~$\tilde\Sigma$. In practice, this means that there exists a coordinate system adapted to a $\Sigma$-foliation for which the reference Riemann tensor can be written as\footnote{Throughout this paper, we denote indices running from 0 to 3 by Greek letters and indices running from 1 to 3 by Roman letters.}
\begin{equation}
	\bar R^\mu{}_{\alpha\nu\beta} = \delta^\mu_a \delta^i_\alpha \delta^b_\nu \delta^j_\beta \,  \bar \CR^a{}_{ibj}(x^k), \label{eq::Riembar_choice}
\end{equation}
where {$\delta^\mu_a$ is the Kronecker symbol} and $\bar \CR^a{}_{ibj}$ is independent of $x^0$ and corresponds to the Riemann tensor of the Riemannian covering space $\tilde\Sigma$.
For instance, in the case $\Sigma$ is a prime manifold %in the sense of Thurston's decomposition of 3-manifolds 
(i.e. $\tilde\Sigma  \in \{\mE^3, \ \mS^3,\ \mH^3, \ \mR\times\mS^2, \ \mR\times\mH^2, \ Nil, \ Sol, \ \tilde{SL}(2,\mR)\}$), then $\bar \CR^a{}_{ibj}$ is related to the standard homogeneous metric of these Riemannian manifolds (they are listed in e.g. \citep[][Section~5.1]{1995_La_Lu}). In particular, in the case of interest for the present paper, for $\tilde\Sigma = \mS^3$ (i.e. the spherical topological class) we have $\bar \CR_{ij} = 2\bar h^{\mS^3}_{ij}$, where $\bar h^{\mS^3}_{ij}$ is a homogeneous and isotropic metric on $\mS^3$, implying
\begin{align}
	\bar R_{\alpha\beta} = 2\, \delta^i_\alpha \delta^j_\beta \;  \bar h^{\mS^3}_{ij}(x^k) \quad {\rm for} \quad \tilde\CM = \mR\times \mS^3. \label{eq_Ricci_S3}
\end{align}
The only case where $\bar \CR^a{}_{ibj} = 0$, and therefore $\bar R_{\mu\nu} = 0$, is for $\tilde\Sigma = \mE^3$ (the Euclidean topological class).

\subsection{The field equations}
\label{sec_Field}

The action of the theory is \citep{2024_Vigneron}
% \begin{align}
% 	S	&\coloneqq \int_\CM\sqrt{-g} \, \left[\frac{1}{2\kappa}\left(\CC^\mu_{\alpha\nu}\CC^\nu_{\beta\mu} - \CC^\mu_{\mu\nu}\CC^\nu_{\alpha\beta}\right) g^{\alpha\beta} + \CL_{\rm m}\right]\dd x^4, \label{eq_action}
% \end{align}
\begin{align}
	S	&\coloneqq \int_\CM\sqrt{-g} \, \left[\frac{1}{2\kappa}\left(R_{\mu\nu} - \bar R_{\mu\nu}\right) g^{\mu\nu} + \CL_{\rm m}\right]\dd x^4, \label{eq_action}
\end{align}
where $\CL_{\rm m}$ is the Lagrangian of matter, $\kappa \coloneqq 8\pi G$. % and $\CC^\mu_{\alpha\beta} \coloneqq \Gamma^\mu_{\alpha\beta} - \bar \Gamma^\mu_{\alpha\beta}$ is called the disformation tensor. 
The field equation resulting from this action is
\begin{equation}
	G_{\alpha\beta} = \kappa\, T_{\alpha\beta} - \Lambda g_{\alpha\beta} + \Top_{\alpha\beta}, \label{eq::ModE1}
\end{equation}
where {we define}
\begin{equation}
	\Top_{\alpha\beta} \coloneqq \bar{R}_{\alpha\beta} - \frac{\bar{R}_{\mu\nu}g^{\mu\nu}}{2} g_{\alpha\beta}. \label{eq::Top_munu}
\end{equation}
Equation~\eqref{eq::ModE1} can be rewritten as
\begin{align}
	R_{\alpha\beta} - \bar{R}_{\alpha\beta} = \kappa \, \left(T_{\alpha\beta} - \frac{T}{2} g_{\alpha\beta} \right) + \Lambda g_{\alpha\beta}, \label{eq:biCoEq}
\end{align}
which highlights more explicitly the difference with general relativity: the metric Ricci tensor is replaced by the difference between this tensor and the reference Ricci tensor.  The action~\eqref{eq_action} also implies that the additional term $\Top_{\alpha\beta}$ in equation~\eqref{eq::ModE1}, compared to the Einstein equation, is conserved {(assuming the matter Lagrangian is independent of $\bar\nabla$)} \citep{2024_Vigneron}:
\begin{align}
	g^{\mu\nu}\left(\nabla_\mu \bar R_{\nu\alpha} - \frac{1}{2} \nabla_\alpha \bar R_{\mu\nu}\right) = 0. \label{eq:biCoCond}
\end{align}
%This is obtained by varying the action with respect to $\bar{\T\nabla}$, assuming the field equation~\eqref{eq:biCoEq} to hold. 
Equation~\eqref{eq:biCoCond}, called the \textit{bi-connection condition}, constraints the additional (diffeomorphism) degrees of freedom, compared to general relativity, coming from the introduction of the reference connection. It does not constrain this connection, but only how it is related to the physical connection.\saut

As presented in the previous section, if a Euclidean spatial topology is chosen, then $\bar R_{\mu\nu} = 0$, implying that equation~\eqref{eq:biCoEq} becomes the standard Einstein equation and that equation~\eqref{eq:biCoCond} becomes trivial. {\it Therefore, GR and {\MyTheory} coincide for this class of topology, and differ for any other type of topology.}\saut 

Equations~\eqref{eq:biCoEq}--\eqref{eq:biCoCond} are equivalent to the ones of the bi-connection theory proposed  by \citet{1980_Rosen}. The only difference (but a fundamental one) is the choice and motivation for the reference connection: Rosen chose a reference connection related to a de Sitter metric in order to remove singularities from general relativity, while in our case, the reference connection is topology dependent as it is related to the universal cover of the spacetime {manifold}~$\CM$. Currently, the main reason for considering {\MyTheory} with the above choice of reference connection is to have a theory admitting a non-relativistic limit in any topology, something not possible with general relativity. {We direct the reader to Section~4.3 in~\citep{2024_Vigneron} for a detailed argumentation. Additionally, the well-posedness of inflation for any background curvature recently came as a second motivation for this theory \citep{2024_Vigneron_et_al_b}.}\saut

An interesting analogy we can make to qualitatively understand {\MyTheory} is to consider the surface of the Earth. The presence of ``matter'', here which we consider to be the mountains, water waves, trees, etc, {intrinsically} curves the surface of the Earth locally.  Still, without this matter, because this surface is topologically a 2-sphere, its {intrinsic} curvature can never be identically zero: it is a topological (thus global) constraint. Therefore, the effect of matter is not to curve the Earth surface, but to induce local deviations from the simplest curvature imposed by topology. This is how we can understand the role of the reference curvature in equation~\eqref{eq:biCoEq}. The interpretation of that equation is that the presence of matter, and/or cosmological constant, induces a departure of the physical spacetime curvature from the ``topological curvature'' $\T{\bar R}$. This must be compared with general relativity, in which matter directly curves spacetime.

\section{The solution}
\label{sec:The_solution}

\subsection{Derivation}

For general relativity, the first exact inhomogeneous solution found was the static vacuum spherically symmetric solution. We are searching for the equivalent solution for {\MyTheory}, in the case $\tilde\Sigma \not= \mE^3$. As said previously, in the case $\tilde\Sigma = \mE^3$, {\MyTheory} is equivalent to GR because the reference curvature is zero. Since we are dealing with closed spatial manifolds, spherical symmetry is only possible if the spatial covering space is $\mS^3$.\footnote{Only two closed 3-manifolds can have spherical symmetry: the 3-sphere $\mS^3$ and the real projective space $\mathbb{R}\mathbb{P}^3$. Both of them are spherical topologies (i.e. $\tilde\Sigma = \mS^3$), therefore $\bar R_{\alpha\beta} = 2 \delta^i_\alpha \delta^j_\beta \bar h^{\mS^3}_{ij}$ for both, by definition of the theory.} Any other choice of closed spatial topology (e.g. the 3-torus for $\tilde\Sigma = \mE^3$ or the Seifert--Weber topology for $\tilde\Sigma = \mH^3$) breaks this symmetry.  Therefore, by definition~\eqref{eq_Ricci_S3}, we can choose a coordinate system $\{\tilde t, \tilde\chi, \theta, \varphi\}$ where the reference Ricci tensor takes the form
 \begin{align}
	\bar R_{\mu\nu}\dd x^\mu \dd x^\nu = 2 \dd\tilde\chi^2 + 2 \sin^2\left(\tilde\chi\right)\dd \Omega^2, \label{eq:Rbar_munu_init}
\end{align}
where $\dd\Omega^2 := \dd\theta^2 + \sin(\theta)^2\dd\varphi^2$. Since we assume spherical symmetry, in the same coordinate system, the line element of the metric can be written, {in full generality}, as
\begin{align}
	\dd s^2 &= \left[-A_1(\tilde\chi,\tilde t) + \beta(\tilde\chi,\tilde t)^2/A_2(\tilde\chi,\tilde t)\right]\dd \tilde t^{\, 2} + 2\beta(\tilde\chi,\tilde t) \dd \tilde t\dd \tilde\chi \nonumber \\
	&\qquad + A_2(\tilde\chi,\tilde t)\dd\tilde\chi^2 + A_3(\tilde\chi,\tilde t) \sin^2(\tilde\chi)\dd\Omega^2. \label{eq:non-static}
\end{align}
From here, we make a first restrictive hypothesis by assuming staticity of the metric in the same coordinate system where~\eqref{eq:Rbar_munu_init} holds. In general relativity, the Birkhoff theorem shows that the assumption of staticity for vacuum spherically symmetric solutions is not a restriction, and corresponds to the general solution. This is not the case in {\MyTheory}, where staticity is expected to be a restrictive assumption. In particular, as will be discussed later, the main consequence of this hypothesis is that the total mass represented by the solution will be zero. In Section~\ref{sec:BH_Alone}, we discuss what we expect to obtain without this hypothesis, and why it will be important as a future prospect to search for a non-static solution.\saut

With staticity, the shift $\beta$ in equation~\eqref{eq:non-static} can be removed by a change of coordinates $\tilde t \mapsto t + f(\tilde\chi)$ without changing the expression of the reference Ricci tensor~\eqref{eq:Rbar_munu_init}, such that we obtain the line element
\begin{align}
	\dd s^2 = -B_1(\tilde\chi)\dd t^2 + B_2(\tilde\chi)\dd\tilde\chi^2 + B_3(\tilde\chi) \sin^2(\tilde\chi)\dd\Omega^2. \label{eq:pingouin}
\end{align}
At this point, we make a second hypothesis which is, a priori, a loss of generality, by assuming
\begin{align}
	B_1(\tilde\chi) = {\rm const}/B_2(\tilde\chi). \label{eq:B1B2=cst}
\end{align}
{This ansatz, inspired by its Schwarzschild counterpart, allows us to obtain an analytical solution.} When solving the system~\eqref{eq:biCoEq}--\eqref{eq:biCoCond}, the following solution is obtained (details can be found in Appendix~\ref{app:Details}):
\begin{align}
	\dd s^2	&= - \frac{\sin(\tilde\chi - \chih/2)}{\sin (\tilde\chi+\chih/2)}\dd t^2 +  \frac{\sin(\tilde\chi+\chih/2)}{\sin(\tilde\chi - \chih/2)} \Rc^2\dd\tilde\chi^2 + \Rc^2 \sin^2(\tilde\chi +\chih/2)\dd \Omega^2, \label{eq:the_solution_har}
\end{align}
where $\chih \in [0,\pi]$ and $\Rc >0$ are free parameters. As will be shown in Section~\ref{sec:Diagram}, the former parameter corresponds to the horizon coordinate of the black hole, and the latter to the curvature radius of the 3-sphere in the case $\chih = 0$. Because the assumption~\eqref{eq:B1B2=cst} is a loss of generality, the above solution might not be the general static vacuum spherically symmetric solution of {\MyTheory}. However, as will be shown in Section~\ref{sec:NR_limit}, the non-relativistic limit of this solution corresponds to the general static vacuum spherically symmetric solution of Newtonian gravitation on $\mS^3$. For this reason, we expect~\eqref{eq:B1B2=cst} not to be a loss of generality and the solution~\eqref{eq:the_solution} to be the general solution.\saut% static vacuum spherically symmetric solution of {\MyTheory}.\saut

With the change of coordinates $\tilde\chi \mapsto \chi - \chih/2$, we obtain the standard line element $\Rc^2 \sin^2(\chi)\dd \Omega$ on $\mS^2$ for the coordinate variables $\theta$ and $\varphi$:
\begin{align}
	\dd s^2	&= - \frac{\sin(\chi - \chih)}{\sin \chi}\dd t^2 +  \frac{\sin\chi}{\sin(\chi - \chih)} \Rc^2\dd\chi^2 + \Rc^2 \sin^2\chi \dd \Omega^2. \label{eq:the_solution}
\end{align}
In these coordinates, the reference Ricci curvature takes the form
 \begin{align}
	\bar R_{\mu\nu}\dd x^\mu \dd x^\nu = 2 \dd\chi^2 + 2 \sin^2\left(\chi -\chih/2\right)\dd \Omega^2.
\end{align}
In the coordinates~\eqref{eq:the_solution_har}, the solution is similar to the Schwarzschild metric in harmonic coordinates, while in the coordinates~\eqref{eq:the_solution}, the solution is similar to the Schwarzschild metric in standard coordinates, with $\frac{\sin(\chi - \chih)}{\sin \chi}$ playing the role of $1-\frac{R_{\rm S}}{r}$. As we will see in the following sections, there are a lot of similarities between our solution and the Schwarzschild metric, with key differences related to topology.

\subsection{Singularities}
\label{sec:singularities}

For $\chih \not=0$, there are two infinite singularities at $\chi = 0$ and $\chi = \pi$, which can be identified through the analysis of the Ricci scalar:\footnote{Contrary to the Schwarzschild metric for which $R_{\mu\nu} = 0$, in our solution, where the vacuum condition rather reads $R_{\mu\nu} = \bar R_{\mu\nu} \not= 0$ for $\mS^3$, it is possible to study true singularities directly via the Ricci scalar instead of the Kretschmann scalar.}
\begin{align}
	R_{\mu\nu} g^{\mu\nu} = \frac{2 + \cos\chih - 3\cos\left(2\chi - \chih\right)}{{\Rc}^2 \sin^2\chi}. \label{eq:RS}
\end{align}
As will be made clear from the Penrose--Carter diagrams of that solution (Figure~\ref{fig_Penrose}), the singularity at $\chi = 0$ is a black hole singularity (BH), i.e. attractive and shielded by a horizon at $\chi = \chih$ which is a coordinate singularity of~\eqref{eq:the_solution}, and the one at $\chi = \pi$ is a {repulsive naked singularity} (RNS), i.e. repulsive and not shielded by a horizon. For this reason we will sometime name our solution, the {``BH-RNS metric''}.\saut

The Ricci scalar~\eqref{eq:RS} shows that there is no $C^2$-extension of the metric at the coordinates $\chi = 0$ and $\chi = \pi$. This is because the Ricci scalar involves second derivatives of the metric \citep[][]{2015_Sbierski}. Furthermore, since the metric is solution of $R_{\mu\nu} = \bar R_{\mu\nu}$, we also have  
\begin{align}
	\bar R_{\mu\nu} g^{\mu\nu} = R_{\mu\nu} g^{\mu\nu} = \frac{2 + \cos\chih - 3\cos\left(2\chi - \chih\right)}{{\Rc}^2 \sin^2\chi}. \label{eq:RSbar}
\end{align}
Because $\bar R_{\mu\nu} g^{\mu\nu}$ involves no derivatives of the metric, the singularities of equation~\eqref{eq:RSbar} shows that there is also no $C^0$-extension of the metric at $\chi = 0$ and $\chi = \pi$.\saut

The fact that we are easily able to determine the $C^0$-inextensibility of the singularities, contrary to general relativity with the Schwarzschild metric for which the task is more complicated \citep{2015_Sbierski}, is because of the presence of an extra structure: the reference curvature. As shown above, it allows us to construct a singular scalar field involving no derivative of the metric.

\subsection{Stable orbits}
\label{sec:Stable_Orbits}

\begin{figure}[t]
	\centering
	\captionsetup{width=.9\linewidth}
	\includegraphics[width=9cm]{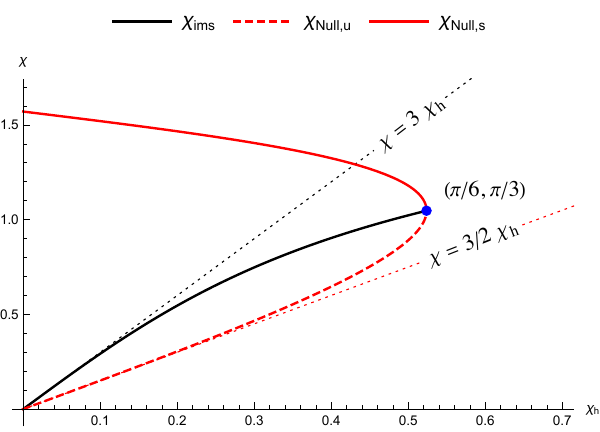}
	\caption{Coordinates of the innermost stable timelike orbit ($\chi_{\rm ims}$), the stable null orbit ($\chi_{\rm Null, s}$) and the unstable null orbit ($\chi_{\rm Null, u}$) as functions of the horizon coordinate $\chih$. These orbits exist only if $\chih <\pi/6$. We also plot in dotted lines the behaviour of the  innermost stable timelike orbit ($\chi_{\rm ims} = 3\chih$) and the unstable Null orbit ($\chi_{\rm ims} = \frac{3}{2}\chih$) for a small horizon radius. In this limit, we retrieve the Schwarzschild result by making the association $\chih \rightarrow R_{\rm S}$.
	\label{fig:Orbits}}
\end{figure}

When restricted to the plane $\theta = \pi/2$, the geodesic equations are
\begin{align}
	&\dot t = k \frac{\sin\chi}{\sin\left(\chi - \chih\right)} \quad;\quad \dot \varphi = \frac{l}{\Rc\sin^2\chi} \quad ;\label{eq:geodesics} \\
	 &\ddot\chi +\frac{\sin\left(\chi-\chih\right)\sin\chih}{2 \Rc^2\sin^3\chi} \, {\dot t}^2 - \frac{\sin\chih}{2\sin\left(\chi-\chih\right)\sin\chi} \, {\dot \chi}^2 - \frac{\sin\left(\chi - \chih\right)}{\cos\chi} \, {\dot\varphi}^2 = 0, \nonumber
\end{align}
where $l$ quantifies the angular momentum of the geodesics, and $k$ its energy.  The first integral $\dot x^\mu \dot x^\nu g_{\mu\nu} = -\epsilon$, where $\epsilon = 1$ for timelike geodesics, and $\epsilon = 0$ for null geodesics, gives
\begin{align}
	\frac{1}{2}\left(\Rc\dot\chi\right)^2 + V_{{\rm eff}, \epsilon}\left(\chi,\chih,l\right) = \frac{k^2-\epsilon}{2},
\end{align}
where
\begin{align}
	V_{{\rm eff},\epsilon}\left(\chi,\chih,l\right) := \frac{\epsilon}{2}\left(\frac{\sin\left(\chi-\chih\right)}{\sin\chi} - 1\right) + \frac{\sin\left(\chi - \chih\right) l^2}{2\sin^3\chi}.
\end{align}
The characterisation of the stable orbits is made by studying the properties of this effective potential. For $\chih < \pi/6$, as for the Schwarzschild metric, there is an innermost stable timelike orbit~$\chi_{\rm ims}$ defined as the radius of the closest minima of $V_{{\rm eff},\epsilon}$ for $\chih$ fixed. It is obtained for ${l = \frac{\sqrt{3}\sin\chih}{\sqrt{2\cos\left(2\chih\right) - 1}}}$ and is
\begin{align}
	\chi_{\rm ims} = \arctan\left[3\tan\chih\right] \quad {\rm for} \quad 0<\chih < \frac{\pi}{6}.
\end{align}
This is consistent with the Schwarzschild result ($r_{\rm ims} = 3R_{\rm S}$) when the horizon radius is small compared to the size of the sphere: $\chi_{\rm ims} = 3\chih + \bigO{\chih^2}$. However, unlike for the Schwarzschild metric, this innermost stable timelike orbit exists only if~$\chih < \pi/6$.
There is also an unstable photon orbit at
\begin{align}
	\chi_{\rm Null, u} = {\rm arccot}\left[\frac{\cos\chih -\sqrt{2 \cos\left(2 \chih\right) -1}}{3 \sin\chih }\right] \quad {\rm for} \quad 0<\chih < \frac{\pi}{6},
\end{align}
which is consistent with the Schwarzschild result ($r_{\rm Null, u} = \frac{3}{2}R_{\rm S}$) since $\chi_{\rm Null, u}  = \frac{3}{2}\chih + \bigO{\chih^2}$. However, unlike for the Schwarzschild metric, there is also a stable photon orbit at
\begin{align}
	\chi_{\rm Null, s}  =  {\rm arccot}\left[\frac{\cos\chih +\sqrt{2 \cos\left(2 \chih\right) -1}}{3 \sin\chih }\right] \quad {\rm for} \quad 0<\chih < \frac{\pi}{6},
\end{align}
which is close to the 3-sphere equator for a small horizon radius, i.e. $\chi_{\rm Null, s}  = \pi/2 + \bigO{\chih}$. Both the stable and unstable photon orbits exist only for $\chih < \frac{\pi}{6}$, as for the innermost stable timelike orbit. The radii $\chi_{\rm ims}$, $\chi_{\rm Null, u}$ and $\chi_{\rm Null, s}$ are represented as functions of $\chih$ in Figure~\ref{fig:Orbits}.\saut

The presence of a stable photon orbit and an upper bound for the horizon radius such that stable causal orbits (i.e. timelike or lightlike) exist can be interpreted as a consequence of the finiteness of the spacelike volume between the BH horizon and the naked singularity (shown in the sections below). For this reason, a more physical solution, i.e. without a repulsive singularity, should also have such an upper bound. The main consequence of these results is that for $\chih > \pi/6$, any geodesic, starting from region I or I' (see Figure~\ref{fig_Penrose}), will fall in the singularity at $\chi=0$ in finite time. However, with finite acceleration, a timelike observer can stay indefinitely in regions~I or I'.

\subsection{Penrose--Carter diagram for $\chih \in ]0,\pi[$}

\begin{figure}[t]
	\centering
	\captionsetup{width=.9\linewidth}
	\includegraphics[width=9cm]{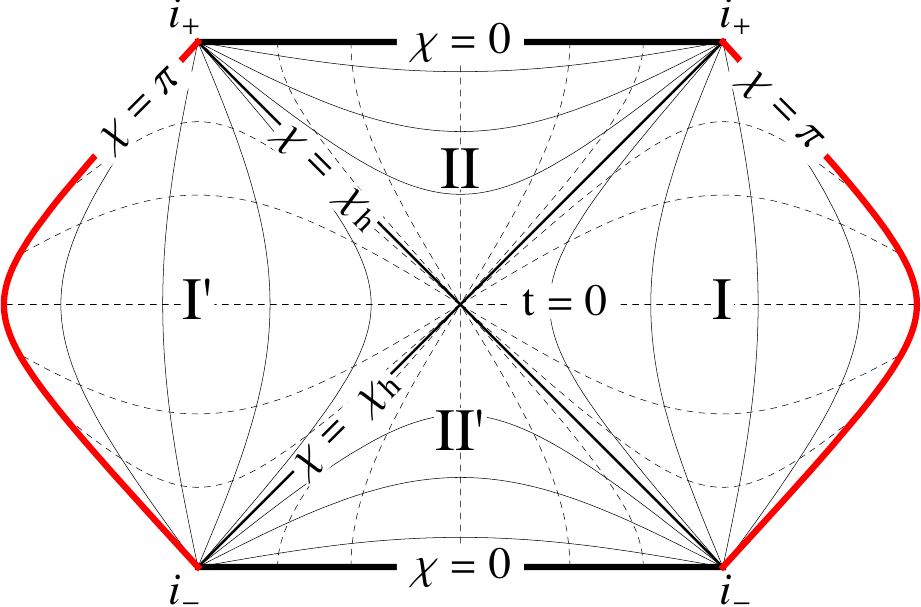}\hfill
	\includegraphics[width=5.8cm]{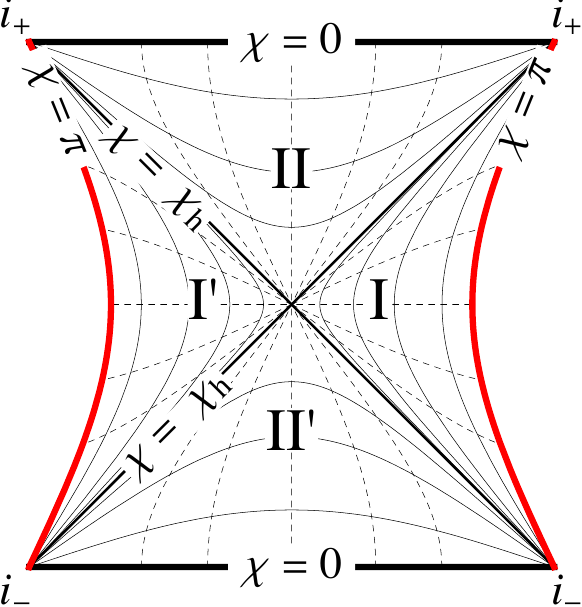}
	\caption{Penrose--Carter diagrams of the maximally extended metric~\eqref{eq:the_solution_KS_compact} with $\Rc = 1$ for~{${0<\chih<\pi/2}$} (left) and for $\pi/2 < \chih < \pi$ (right). Iso-$\chi$ lines are full, and iso-$t$ lines are dashed. The {repulsive naked singularities} correspond to the red thick lines, while the black hole singularities correspond to the black thick lines. {The causal properties of these two diagrams are the same. Their difference lies in the relative spatial volume between regions I (or I') and II (or II'), that is controlled by $\chih$ for a fixed $\Rc$.}
\label{fig_Penrose}}
\end{figure}

When solving the geodesic equations~\eqref{eq:geodesics}, the radial null geodesics are given by
\begin{align}
	t = \pm \Rc\cos\left(\chih\right)\left[\chi + \tan\left(\chih\right)\log\left|\sin\left(\chi-\chih\right)\right|\right] + {\rm const}.
\end{align}
Then, applying the same strategy as for regularising the horizon of a Schwarzschild metric, we can obtain Kruskal--Szekeres like coordinates from the line element~\eqref{eq:the_solution}, with the change of coordinates:
\begin{align}
	&{\rm for \ \chi > \chih} \quad 
	\begin{dcases}
		\tilde v &:= \sqrt{\frac{\sin(\chi-\chih)}{\sin\chih}}\exp\left({\frac{\chi}{2\tan\chih}}\right)\sinh\left(\frac{t}{2\Rc\sin{\chih}}\right) \\
		\tilde u &:= \sqrt{\frac{\sin(\chi-\chih)}{\sin\chih}}\exp\left({\frac{\chi}{2\tan\chih}}\right)\cosh\left(\frac{t}{2\Rc\sin{\chih}}\right)
	\end{dcases},\\
	&{\rm for \ \chi < \chih} \quad 
	\begin{dcases}
		\tilde v &:= \sqrt{-\frac{\sin(\chi-\chih)}{\sin\chih}}\exp\left({\frac{\chi}{2\tan\chih}}\right)\cosh\left(\frac{t}{2\Rc\sin{\chih}}\right) \\
		\tilde u &:= \sqrt{-\frac{\sin(\chi-\chih)}{\sin\chih}}\exp\left({\frac{\chi}{2\tan\chih}}\right)\sinh\left(\frac{t}{2\Rc\sin{\chih}}\right)
	\end{dcases},
\end{align}
leading to the line element
\begin{align}
	\dd s^2	&= \frac{4 \Rc^2 \sin^3\chih}{\sin\chi} e^{-\chi \cot\chih}\left(-\dd \tilde v^2 +  \dd \tilde u^2\right) + \Rc^2 \sin^2\chi \, \dd \Omega, \label{eq:the_solution_KS}
\end{align}
where $\tilde u$ and $\tilde v$ take values in $\mathbb{R}$. These coordinates can be compactified by the usual change of coordinates
\begin{align}
	 v &:= \arctan\left(\tilde u+\tilde v\right) + \arctan\left(\tilde u-\tilde v\right), \\
	 u &:= \arctan\left(\tilde u+\tilde v\right) - \arctan\left(\tilde u-\tilde v\right),
\end{align}
which leads to the line element
\begin{align}
	\dd s^2	&= \frac{\Rc^2 \sin\chih}{\sin\chi} e^{-\chi \cot\chih}\frac{-\dd  v^2 +  \dd  u^2}{\cos^2\left(\frac{ v+  u}{2}\right)\cos^2\left(\frac{ v -  u}{2}\right)}  + \Rc^2 \sin^2\chi \, \dd \Omega. \label{eq:the_solution_KS_compact}
\end{align}
This line element is regular at the horizon, and because no $C^0$-extension of the metric can be made at the singularities $\chi = 0$ and $\chi = \pi$ (as argued in Section~\ref{sec:singularities}), it represents the maximal extension of~\eqref{eq:the_solution}.\saut

\begin{figure}[t]
	\centering
	\captionsetup{width=.9\linewidth}
	\includegraphics[width=8cm]{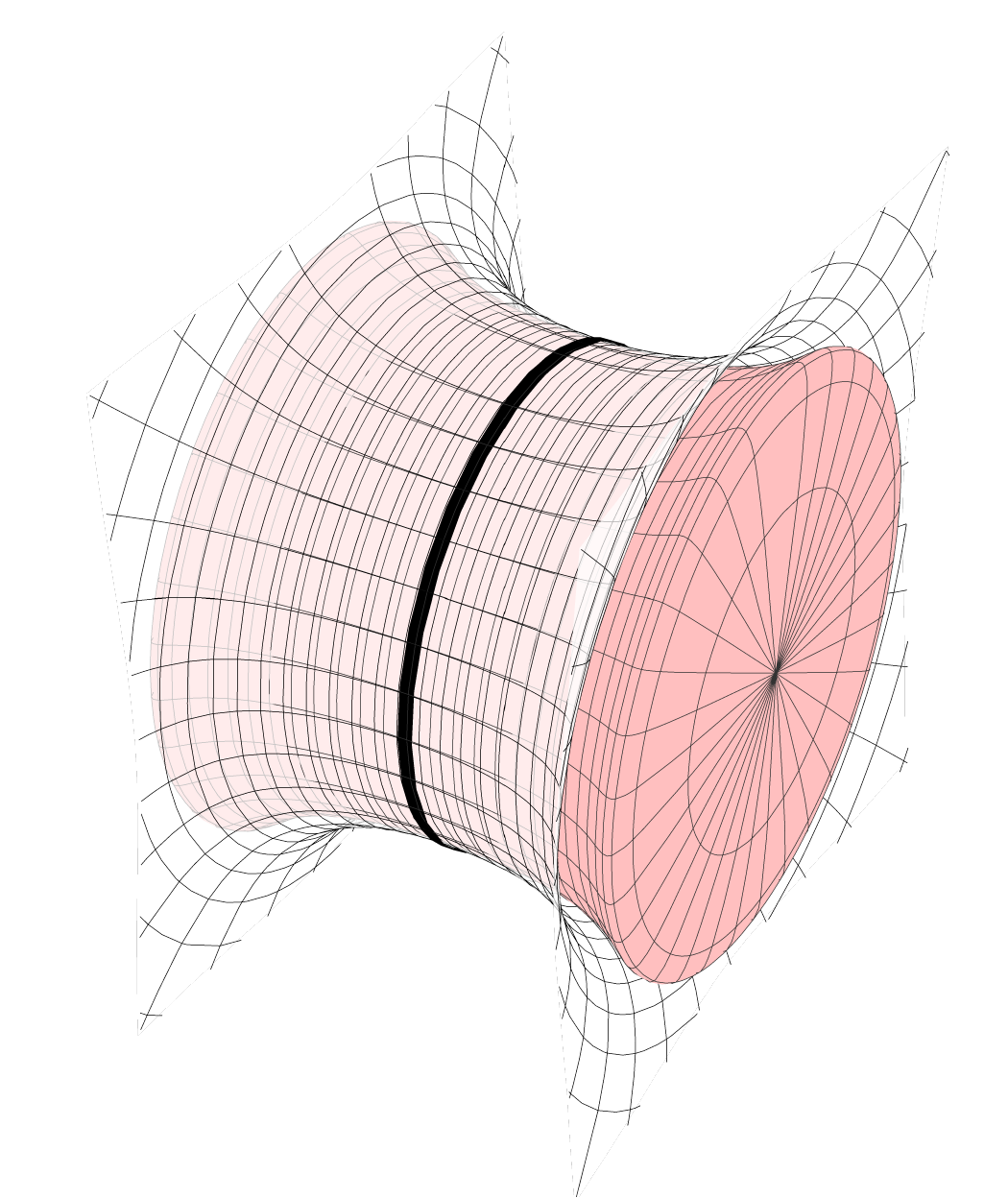}
	\caption{2-dimensional section (with $\theta = \pi/2$ and $t=0$) of the spacelike hypersurface of the Penrose--Carter diagram for the Schwarzschild metric (translucent white) and our solution for $0 < \chih < \pi$ (red). Compact isotropic coordinates are used to draw this figure, as detailed in Appendix~\ref{sec:Iso_Coord}. The black hole horizon is represented by the black thick line, regions I and I' are respectively on the right and the left of this line. Our solution is closed in place of the spatial infinities of Schwarzschild. Template of the figure inspired from \citep{2014_Clifton}.\label{fig_Lattice_double}}
\end{figure}

From~\eqref{eq:the_solution_KS_compact}, we can draw the Penrose--Carter diagram of our solution, shown in Figure~\ref{fig_Penrose} for $0 < \chih < \pi/2$ (left panel) and for $ \pi/2 < \chih < \pi$ (right panel). {These diagrams are similar to the Schwarzschild diagram but with the difference that the spatial infinities are replaced by repulsive naked singularities at finite distance.} This implies that the twin regions~I and~I' have a finite spacelike volume. Therefore, while the metric cannot be extended at $\chi = \pi$, the induced spacelike distance (or volume) can, which is the reason why we can interpret this solution as describing an $\mS^3$ universe, as argued in more details in Section~\ref{sec:Cond_Topo}. This is represented in Figure~\ref{fig_Lattice_double} where we draw a 2-dimensional section of the spacelike hypersurface of the Penrose--Carter diagram at $t=0$ for the Schwarzschild metric (transparent white) and our solution (red). Our solution is similar to Schwarzschild close to the horizon, but the usual flat infinities are closed on both sides. This shows that the topology obtained a posteriori, i.e. after solving the field equations, is indeed the same as the one chosen a priori with the choice~\eqref{eq_Ricci_S3}.\saut

{The repulsive naked singularities at $\chi = \pi$ behave like the Schwarzschild naked singularity (i.e. Schwarzschild metric with negative mass parameter, see e.g. \citep{2006_Gleiser_et_al}), as can be seen in the limit $\chi \rightarrow \pi$ and $\chih \rightarrow 0$ where the metric~\eqref{eq:the_solution} reduces to the Schwarzschild metric (associating $\pi - \chi$ to $r$) with negative mass. Therefore, it is possible for an accelerating timelike observer to stay infinitely close to $\chi = \pi$. But reaching $\chi = \pi$ would require infinite acceleration or energy (i.e. $k\rightarrow\infty$) for such an observer. However, null geodesics can reach $\chi = \pi$ in finite affine parameter as can be seen from the diagrams in Figure~\ref{fig_Penrose}. This is not possible for the repulsive singularity at $\chi = 0$ in {region} II' {of those diagrams}, from which any timelike observer will necessarily cross the white hole horizon delimiting {region} II' in finite proper time.}\saut

\subsection{Penrose--Carter diagrams for $\chih \in \{0,\pi\}$}
\label{sec:Diagram}

\begin{figure}[t]
	\centering
	\captionsetup{width=.9\linewidth}
	\begin{subfigure}{0.49\linewidth}
		\includegraphics[width=7cm]{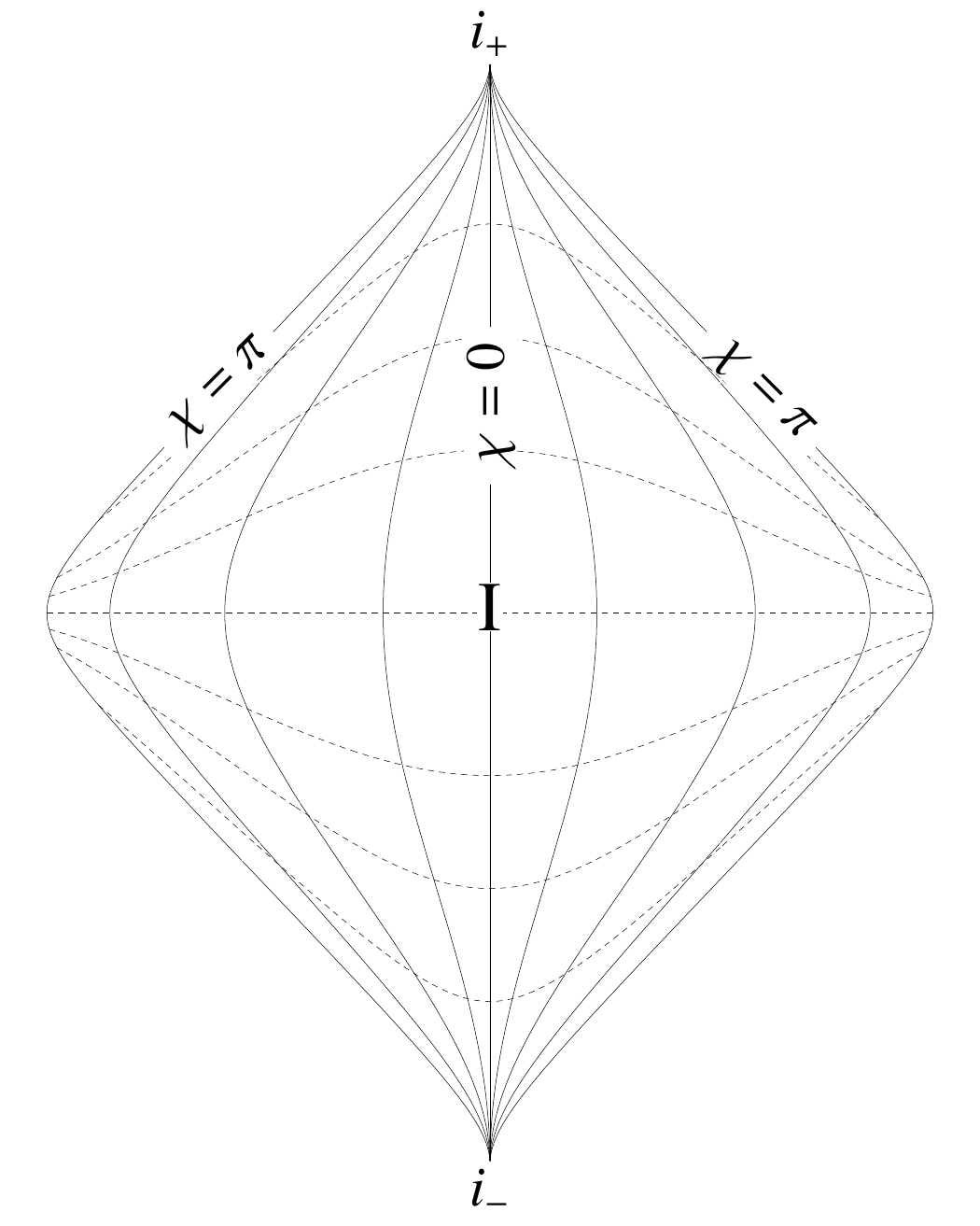}
	\end{subfigure}
	\begin{subfigure}{0.49\linewidth}
		\includegraphics[width=8cm]{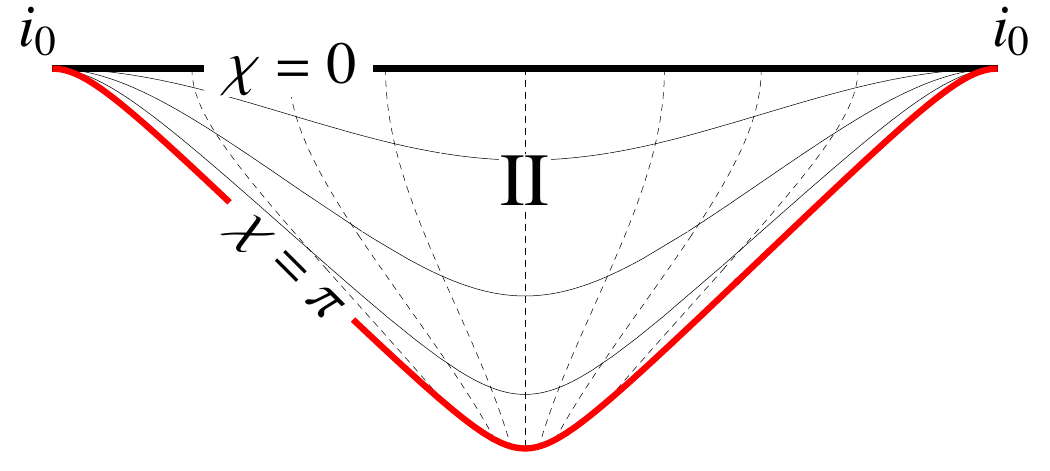}
	\end{subfigure}
	\caption{(Left) Penrose--Carter diagram for a vacuum $\mS^3$ universe, as described by the metric~\eqref{eq_vide} with $\chih = 0$. {Each point on the diagram corresponds to $\mS^2$, apart for points at $\chi = \pi$ and points at $\chi=0$ each corresponding to one of the two (arbitrary) poles of the 3-sphere. Accordingly, spatial sections are closed in this diagram.} %{Qualitatively}, it corresponds to a subset of the Minkowski diagram, from which the two spacelike infinities are cut and glued together at $\chi = \pi$. It is also a subset of region I in Figure~\ref{fig_Penrose}.
	(Right) Penrose--Carter diagram as described by the metric~\eqref{eq_chelou} with $\chih = \pi$. %This diagram is a subset of the Minkowski diagram, from which the two timelike infinities are cut and replaced by two singularities. It is also a subset of region II in Figure~\ref{fig_Penrose}.
	\label{fig_Penrose_0_Pi}}
\end{figure}

The {repulsive naked singularity} line tends to the Schwarzschild spatial infinity when {${\chih \rightarrow 0}$}, but it never reaches it. Instead, for $\chih = 0$, the line element becomes
\begin{align}
	\dd s^2	&= -\dd t^2 + \Rc^2\left(\dd\chi^2 +  \sin^2\chi \,\dd \Omega^2\right). \label{eq_vide}
\end{align}
%Its Penrose--Carter diagram can be seen, qualitatively, as a subset of the Minkowski diagram from which the spacelike infinities are cut and glued together (left of Figure~\ref{fig_Penrose_0_Pi}), and no singularities are present anymore (the Ricci scalar~\eqref{eq:RS} is fully regular for $\chi \in [0,\pi]$). {(This is only a qualitative view, since the metric is not the same.)} 
This metric represents a universe with an $\mathbb R\times\mS^3$ spacetime topology, whose total mass is zero, as is made clear from the Komar mass calculated in Section~\ref{sec:Mass}, and in which no singularities are present anymore. Such an empty (we recall that in the context of {\MyTheory} ``empty'' means $R_{\mu\nu} = \bar R_{\mu\nu}$) static homogeneous $\mS^3$ universe is not possible in general relativity because the presence of spatial curvature in the expansion law prevents this solution to exist. As shown in \citep{2023_Vigneron_et_al_b}, this is not the case anymore with {\MyTheory} in which the expansion law does not feature anymore the spatial curvature, hence allowing for an empty static homogeneous $\mS^3$ universe to be solution. This is further discussed in Section~\ref{sec:Why}.\saut

When $\chih \rightarrow \pi$, the {repulsive naked singularity} tends to the BH horizon but, again, never reaches~it. Instead, for $\chih = \pi$, {the diagram is a subset of region II in Figure~\ref{fig_Penrose} {in} which an initial space-like singularity is introduced.} 
%of the Minkowski diagram (right of Figure~\ref{fig_Penrose_0_Pi}), from which the timelike infinities are cut and replaced by singularities (the Ricci scalar~\eqref{eq:RS} still diverges for $\chi \in \{0,\pi\}$). 
This case is really peculiar, since the spacelike hypersurfaces are infinite, and any timelike or lightlike curve begins at $\chi = \pi$ and ends up at $\chi = 0$ in finite affine parameter. With an argument similar to the one developed in Section~\ref{sec:Cond_Topo}, this compactness of the time threading still allows us to interpret the topology described by the solution with $\chih=\pi$ to be $\mathbb R\times \mS^3$. The reason why the spacelike hypersurfaces are infinite is because they have been so much tilted in $\mR\times\mS^3$ when taking $\chih\rightarrow \pi$, that the $\mR$-direction, which is timelike for $\chih<\pi$, now lies within these spacelike hypersurfaces, and their orthogonal direction, i.e. the time threading, now lies in $\mS^3$. This can be seen with the BH-RNS metric which, for $\chih = \pi$, is
\begin{align}
	\dd s^2	&= \dd t^2 + \Rc^2\left(-\dd\chi^2 + \sin^2\chi \,\dd \Omega^2\right). \label{eq_chelou}
\end{align}
The roles played by $t$ and $\chi$ are inverted, but, still, $\chi$, $\theta$, and $\varphi$ are coordinates on $\mS^3$. {Unless otherwise specified, this peculiar case will be excluded in the following discussions.}

\subsection{Mass}
\label{sec:Mass}

The notions of mass or energy in general relativity often require the hypothesis of asymptotic flatness (or at least asymptotical conicity, see e.g. Definition 1.11 in \citep{2022_Kroencke_et_al}), and/or the introduction of a reference flat metric (e.g. the ADM energy \citep{2008_ADM} or the Brown--York energy \citep{1993_Brown_et_al}). Such definitions cannot be used to determine the mass in our solution, where the volume of the spacelike hypersurface between the BH horizon at $\chi = \chih$ and the {RNS} at $\chi = \pi$ is finite. One notion of mass not directly related to asymptotic flatness is the Komar mass. It is defined for spacetimes having a timelike Killing vector $k^\mu$. Inside a spacelike surface~$\CS_t \sim \mS^2$ on a hypersurface $\Sigma_t$, the mass is given by~\citep{2012_GG}
\begin{align}
	M_{\rm K} \coloneqq -\frac{1}{\kappa}\int_{\CS_t} \left(s_\mu n_\nu - n_\mu s_\nu\right)\nabla^\mu k ^\nu\sqrt{q} \, \dd^2 y, \label{eq:MK}
\end{align}
where $n^\mu$ is the normal unit vector to $\Sigma_t$, $s^\mu$ is the normal unit vector to $\CS_t$ within $\Sigma_t$ and oriented outwards from $\CS_t$, and $q = {\rm det} (q_{\mu\nu})$ where $q_{\mu\nu}$ is the metric on $\CS_t$ induced by $g_{\mu\nu}$.\saut

The timelike Killing vector $k^\mu$ is defined up to a constant factor. For asymptotically flat spacetimes, this factor is chosen such that $k^\mu = (1,0,0,0)$ at infinity. So, the Komar mass is actually not uniquely defined and, as for the ADM mass, it requires a reference observer at infinity to make the definition unique. Nevertheless, the definition \eqref{eq:MK} does not explicitly require an infinite volume (and flatness), and therefore can be considered for a metric inducing a finite volume, for which the mass will necessarily (unless there is a natural choice of lapse) be defined up to a constant factor.\saut

We consider our solution written in the form~\eqref{eq:the_solution}.  The vector $(1,0,0,0)$ is a timelike Killing vector for $\chi > \chih$. We choose $\CS_t$ such that it encompasses the BH horizon. The definition~\eqref{eq:MK} leads to
\begin{align}
	M_{\rm K}^{\rm BH} = \frac{\Rc}{2}\sin\chih. \label{eq:M_K^BH}
\end{align}
As expected, $M_{\rm K}$ does not depend on the radius $\chi$ of the surface $\CS_t$, since $\CS_t$ is located in vacuum for any value of $\chi > \chih$. This is a known property of the Komar mass in GR, which also holds in {\MyTheory} as long as either $k^\mu$ or the normal vector $n^\mu$ to $\Sigma_t$ lie in the kernel of $\bar R_{\mu\nu}$ (see Appendix~\ref{app_Komar}).\saut

The mass \eqref{eq:M_K^BH} is the mass of the BH. By reversing the orientation of the surface~$\CS_t$, we directly obtain the mass $M_{\rm K}^{\rm RNS}$ of the repulsive naked singularity as the opposite of the one of the BH: ${M_{\rm K}^{\rm RNS} = -M_{\rm K}^{\rm BH}}$. Therefore the total mass in the static region I of the Penrose--Carter diagram (equivalently, in the static region I') is zero. This does not depend on the normalisation of the timelike Killing vector in the definition of the Komar mass.  This result is discussed in Section~\ref{sec:Why}.

\subsection{{Which topology?}}
\label{sec:Cond_Topo}

\begin{figure}[t]
	\centering
	\captionsetup{width=.9\linewidth}
	\includegraphics[width=7cm]{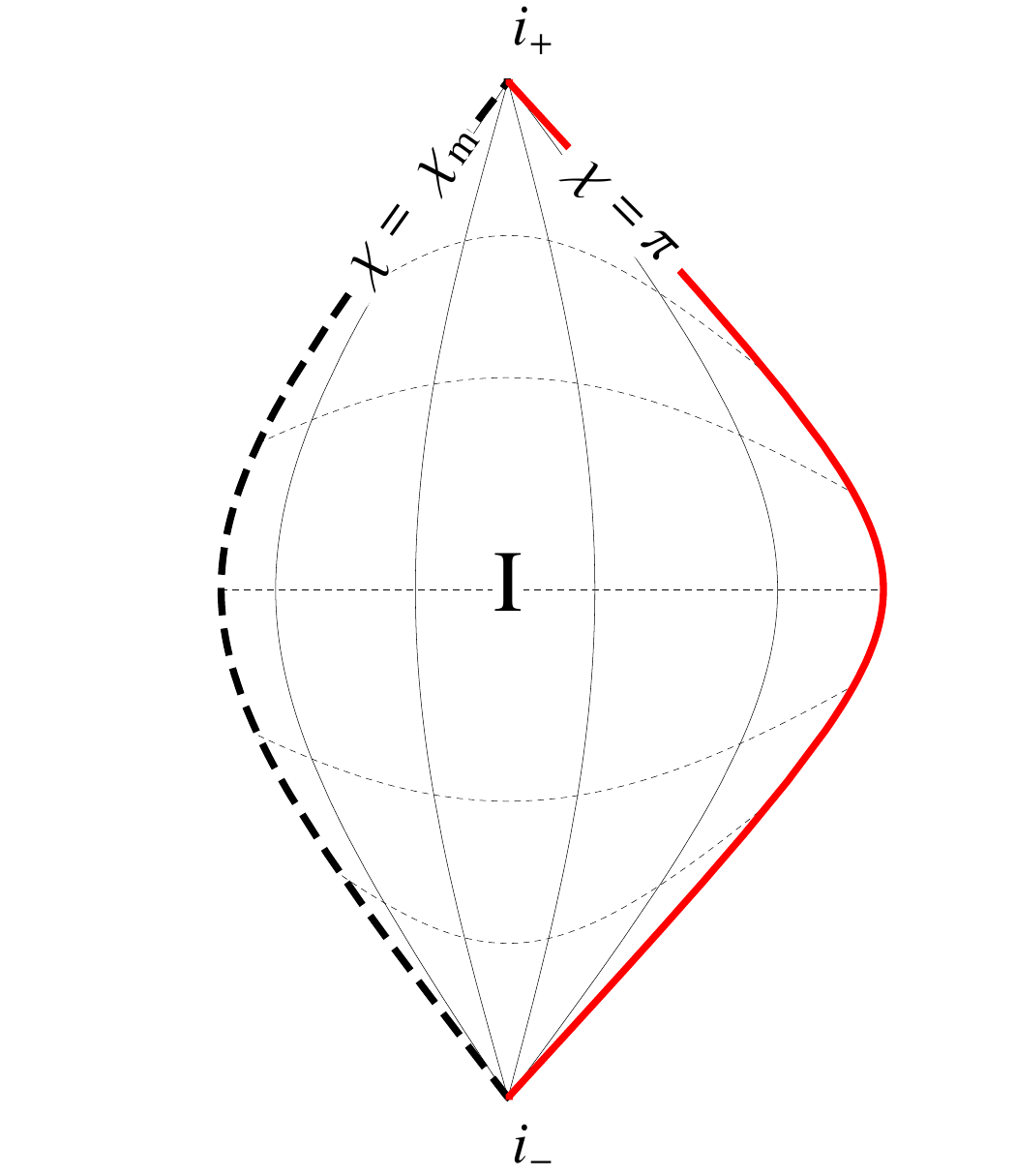}
	\caption{Penrose--Carter diagram for~{${\chih=\pi/4}$} and $\Rc = 1$, in which a spherically symmetric {\it static} compact distribution of matter encompasses the BH horizon. The surface of this matter is defined to be at $\chi_{\rm m} > \chih$ and is represented as a black thick dashed line. {We expect that there should be an upper bound on the mass of the compact object (for a fixed radius~$\Rc$) for this static (non-vacuum) configuration to be possible.}\label{fig_PenroseStar}}
\end{figure}

%{\subsubsection{Finite volume}}

The metric is singular and not continuously extensible at $\chi =0$ and $\chi = \pi$. Considering either the full extension of the metric [equation~\eqref{eq:the_solution_KS_compact}], i.e. describing regions I, II, I' and II', or only part of this extension, e.g. describing only regions I and II [line element~\eqref{eq:the_solution}], or only regions I and I' [line element~\eqref{eq:iso_BHWH}], the manifold on which is defined the metric, as a smooth continuous tensor field, is always $\mR^2\times\mS^2$. On the other hand, the reference Ricci curvature $\bar R_{\mu\nu}$ is defined, as a smooth continuous tensor field, on $\mR\times\mS^3$.

This situation is similar to the solution of Newtonian gravitation for a point mass in $\mR^3$. In this case, we start with a (flat) spatial metric defined everywhere on $\mR^3$ and solve the Poisson equation in vacuum with spherical symmetry. The solution for the gravitational potential turns out to be defined on $\mR\times\mS^2 \simeq \mR^3\setminus\{0\}$. However, we do not consider the spatial topology to be $\mR\times\mS^2$: we still consider it to be $\mR^3$. Can we do the same thing with respect to the {BH-RNS} metric? In other words: while mathematically $g_{\mu\nu}$ is defined on $\mR^2\times\mS^2$, does it make sense, physically, to interpret the topology it describes to be rather $\mR\times\mS^3$? {In the following, we shall argue {in favour of} this interpretation.}\saut

First, having $\bar R_{\mu\nu}$ defined on $\mR\times\mS^3$ and $g_{\mu\nu}$ defined on $\mR^2\times\mS^2$ used together in the same equation [the {\MyTheory} equations~\eqref{eq:biCoEq}--\eqref{eq:biCoCond}] is not in contradiction. Indeed, $\mR^2\times\mS^2$ can be (densely) embedded in $\mR\times\mS^3$ (i.e. $\mR^2\times\mS^2$ corresponds to $\mR\times\mS^3$ from which two lines have been removed) which implies that $g_{\mu\nu}$ can be seen as being defined on a subset of $\mR\times\mS^3$, i.e. everywhere away from the singularities.\saut

Second, to our opinion, the key property for {interpreting} the topology described by the {BH-RNS} metric to be $\mR\times\mS^3$ rather than $\mR^2\times\mS^2$, is to have finiteness of the spacelike hypersurface volume as induced by the metric: i.e. on a spacelike foliation $\{\Sigma_t \sim \mathbb R\times\mS^2\}_{t\in\mathbb R}$ of $\mR^2\times\mS^2$, the volume of the hypersurfaces of this foliation should be finite: 
 \begin{align}
	\int_{\Sigma_t} \sqrt{{\rm det} h_{ij}}\,\dd^3 x < \infty, \label{eq:cond_Topo}
\end{align}
where $h_{ij}$ is the metric induced by $g_{\mu\nu}$ on $\Sigma_t$. \saut

Property~\eqref{eq:cond_Topo} can never be fulfilled by the Schwarzschild metric regardless of the foliation we consider, due to the spatial infinities. Therefore, while this metric is also defined on $\mR^2\times\mS^2$ which can be embedded in $\mR\times\mS^3$, it makes no sense, with respect to condition~\eqref{eq:cond_Topo}, to interpret the Schwarzschild metric as describing a spherical universe. The same applies for the Minkowski metric defined on $\mR^4$ which can also be embedded in $\mR\times\mS^3$, i.e. a point is added at spatial infinity.\saut

The situation is different with the {BH-RNS} metric for which the spacelike volume around each {repulsive naked singularity} is finite, and property~\eqref{eq:cond_Topo} is fulfilled. This can be seen with the Kruskal--Szekeres line element~\eqref{eq:the_solution_KS_compact}, or more easily with the isotropic line element~\eqref{eq:iso_BHWH} describing regions I and I' (derived in Appendix~\ref{sec:Iso_Coord}). The finiteness of the spacelike volume is schematized in Figure~\ref{fig_Lattice_double}.\saut

To summarise, the finiteness of the spacelike hypersurface volume and the fact that $\mathbb R^2\times\mS^2$ is a (dense) subset of $\mathbb R\times\mS^3$ are the reasons why we can \textit{interpret} our solution as describing a $\mathbb{R}\times\mS^3$ universe. That interpretation is further supported by the fact that we can remove the horizon and fill the BH with a finite volume {\it static} distribution of matter, letting only the repulsive naked singularity, around which the volume is spacelike and finite. This is represented in Figure~\ref{fig_PenroseStar}. Overall, the present discussion highlights the way solutions should be searched for in {\MyTheory}:
\begin{enumerate}[label=(\roman*)]
	\item a spacetime topology must be chosen. This sets the reference curvature $\bar R_{\mu\nu}$, as described in Section~\ref{sec_Rbar}.
	\item a Lorentzian (physical) metric solution of~\eqref{eq:biCoEq} is considered. That solution, {\it a posteriori}, might be only defined on a (dense) subset of the initial topology, as is the case with the {BH-RNS} metric.
\end{enumerate}
Note that this would have been contradictory to find an infinite volume as induced by the metric solution of the {\MyTheory} equations \eqref{eq:biCoEq}--\eqref{eq:biCoCond}, while by hypothesis and definition of the theory, we initially considered $\mathbb{R}\times\mS^3$ and defined everywhere on this manifold the Ricci tensor $\T{\bar R}$. Therefore, we suspect that any solution of \eqref{eq:biCoEq}--\eqref{eq:biCoCond} will necessarily have a finite spacelike hypersurface volume (with or without singularities) once we consider spherical topologies, i.e. once we take $\bar R_{\mu\nu} = 2\,\delta_\mu^i \delta_\nu^j \, \bar h^{\mS^3}_{ij}$.\saut

\begin{remark}
{Due to the left-right symmetry of the diagrams~\ref{fig_Penrose}, a second possibility for the spatial topology of the maximally extended solution is the real projective space $\mathbb{R}\mathbb{P}^3$, i.e. the 3-sphere for which {antipodal points are identified}. This argument is equivalent to the one used for the de Sitter metric which can describe either $\mathbb{R}\times\mathbb{S}^3$ or $\mathbb{R}\times\mathbb{R}\mathbb{P}^3$ \citep[e.g.][]{1999_Louko_et_al, 2003_McInnes, 2017_Ong_et_al}. To represent this solution, the left-half of the Penrose-Carter diagram is removed and replaced by a vertical line {along which each point is a} real projective plane $\mathbb R\mathbb P^2$. %(see Figure~\ref{}). 
This procedure is described for the de Sitter space in Section~2 of \citep{2003_McInnes}.}
\end{remark}

\section{Discussions}
\label{sec:Discussions}

\subsection{Compatibility with Newtonian gravitation on $\mS^3$}
\label{sec:NR_limit}

As shown in \citep{2024_Vigneron}, general relativity is not compatible with Newtonian gravitation in non-Euclidean spatial topologies, i.e. for which the covering space is not $\mE^3$. Currently, the main motivation for the construction of {\MyTheory} is to have a relativistic theory admitting such a compatibility, i.e. having a non-relativistic limit in any topology. In this section, we check the compatibility of our solution with the static vacuum spherically symmetric solution obtained in the Newtonian theory on $\mS^3$, which was defined in \citep{2022_Vigneron_b, 2023_Vigneron_et_al_a}.\saut

Newtonian gravitation on $\mS^3$ is described by similar equations as Newtonian gravitation on a Euclidean topology, i.e. classical Newtonian theory \citep[see][for a summary of the full system of equations]{2023_Vigneron_et_al_a}. In particular, the gravitational potential $\Phi$ is given by the cosmological Poisson equation
\begin{align}
	\Delta_{\mS^3} \Phi = \frac{\kappa}{2}\left(\rho - M_{\rm tot}/V_\Sigma \right), \label{eq:Poisson}
\end{align}
where $V_\Sigma \propto a(t)^3$ is the volume of $\mS^3$ and $M_{\rm tot}$ is the total (Newtonian) mass in~$\mS^3$. The only difference with the Euclidean Newtonian theory is the spatial metric which is curved and has the form $h_{ij}(t,x^k) = a^2(t) \bar h^{\mS^3}_{ij}(x^k)$.\footnote{Actually, the most general form of the theory features a metric with anisotropic expansion. However, this does not change the discussion of this section since we consider a static spacetime.} The scale factor $a(t)$ is solution to the Friedmann equation ${3\ddot{a}/a = -4\pi GM_{\rm tot}/V_\Sigma}$ (we assumed $\Lambda = 0$). For this reason, a static solution, for which $\dot{a} = 0$, of Newtonian gravitation on $\mS^3$ necessarily requires $M_{\rm tot} = 0$.\footnote{This results actually holds for any closed topology, and therefore holds in the classical Newtonian theory, since the same Friedmann equation is valid.}\saut

The spherically symmetric static vacuum (i.e. with $M_{\rm tot} = 0$) solution of \eqref{eq:Poisson} is obtained by taking $\rho = M_{\rm N} \left[\delta^{\mS^3}(\chi) - \delta^{\mS^3}(\pi-\chi)\right]$, where $\delta^{\mS^3}$ is the Dirac distribution on~$\left(\mS^3,\T{\bar h}^{\mS^3}\right)$ and $M_{\rm N}$ is the Newtonian mass of the point masses. The solution is
\begin{align}
	\Phi = \frac{M_{\rm N}}{\Rc}\cot\chi,
\end{align}
where $\Rc$ is the curvature radius of the 3-sphere. As for the relativistic solution~\eqref{eq:the_solution}, it represents an attractive and a repulsive point masses of opposite Newtonian mass $\pm M_{\rm N}$ at opposite poles of $\mS^3$, taken to be of radius $\Rc$.\saut

The dictionary between the metric solution of {\MyTheory} and Newtonian gravitation on~$\mS^3$ was derived in \citep[][Section 6.3]{2024_Vigneron} and is similar to the standard dictionary in general relativity. In particular in a gauge without shift, as is the case with the line element~\eqref{eq:the_solution}, we have $g_{00} = -c^2 + 2\Phi + \bigO{1/c^2}$. By reintroducing $c^2$ in~\eqref{eq:the_solution} as follows: $t \rightarrow c\,t$ and $\chih \rightarrow \arcsin\left[2M^{\rm BH}_{\rm K}/(c^2\,\Rc)\right]$, we obtain  $g_{00} = -c^2 + 2\frac{M^{\rm BH}_{\rm K}}{\Rc}\cot\chi + \bigO{1/c^2}$, consistent with the Newtonian potential derived above, associating $M^{\rm BH}_{\rm K}$ to $M_{\rm N}$. In conclusion, the {BH-RNS} solution is the relativistic equivalent of the Newtonian solution featuring a positive and negative equal point masses at opposite poles of $\mS^3$.

\subsection{Total mass and staticity}% $M_{\rm K}^{\rm tot} = 0$?}
\label{sec:Why}

Just like it could have been expected from the Newtonian limit presented in the previous section, having $M_{\rm K}^{\rm tot} = 0$ was also expected from the homogeneous and isotropic solution of {\MyTheory}. As shown in \citep{2023_Vigneron_et_al_b}, the expansion laws for such a solution do not depend anymore on the spatial curvature:
\begin{align}
	\begin{dcases}
		3H^2 &= \kappa \bar\rho + \Lambda, \quad \forall \Omega_K\\
		3\frac{\ddot a}{a} &= -\frac{\kappa}{2} (\bar\rho + 3\bar p) + \Lambda,
	\end{dcases}\label{eq:expansion_laws}
\end{align}
where $\bar\rho$ and $\bar p$ are homogeneous density and pressure, and $\Omega_K$ is the curvature parameter. In the case $\bar\rho > 0$ and $\bar p = 0$, because $\Omega_K$ is not present anymore in the first expansion law, no static solution on $\mS^3$ is possible, for any value of $\Lambda$. This differs from general relativity for which such a solution exists in the form of the Einstein static model. The only possibility for a static homogeneous solution on $\mS^3$ in {\MyTheory} is
\begin{align}
	%\left(\bar\rho = \bar p = \Lambda/\kappa = 0\right) \qquad  {\rm or} \qquad 
 \Lambda = -\kappa\bar\rho \quad; \quad \bar\rho = - \bar p,
\end{align}
which implies $R_{\mu\nu} = \bar R_{\mu\nu}$. In the case $\Lambda = 0$, a zero homogeneous density is required. If we extrapolate this result to inhomogeneous solutions and replace $\bar\rho$ and $\bar p$ by the spatial average density $\average{\rho}{\Sigma_t} = M_{\rm tot}/V_\Sigma$ and pressure $\average{p}{\Sigma_t}$ on $\Sigma_t$, this implies that any static solution on $\mS^3$ must have a zero total mass; and reversely, any inhomogeneous solution with $\average{\rho}{\Sigma_t} > 0$ and $\Lambda = 0$ will necessarily be non-static. This gives a qualitative justification of the zero mass result obtained from the solution~\eqref{eq:the_solution}.\saut

Of course the conclusion drawn from the extrapolation from a homogeneous solution to an inhomogeneous one is far from being fully justified, especially since additional kinematic terms are expected to be present in the average expansion laws, like they are in GR~\citep{Back_I}. A full justification requires a careful analysis of the averaging problem in differential geometry, in the case of {\MyTheory}. Nevertheless, as for the non-relativistic limit of GR (in Euclidean topologies) \citep{1997_Buchert_et_al}, this extrapolation holds in the non-relativistic limit of {\MyTheory}, for which the laws~\eqref{eq:expansion_laws}, with $\average{\rho}{\Sigma}$ and $\average{p}{\Sigma}$ in place of $\bar\rho$  and $\bar p$, hold regardless of the inhomogeneities present in $\Sigma$,  for $\rho$ being smooth as well as for $\rho$ being, e.g., a Dirac comb \citep{2023_Vigneron_et_al_b, 2023_Vigneron_et_al_a}. This gives us confidence in our prediction that a solution of {\MyTheory} with positive mass cannot be static.\saut

Note that, since {\MyTheory} is equivalent to GR in Euclidean (closed) topologies, this also suggests that closed static solutions of GR should necessarily have a zero total mass. Using a quasi-local definition for the mass, i.e. involving a surface integral as for the Komar mass~\eqref{eq:MK}, this property for GR is exact in closed topologies. Indeed, evaluating the mass on both (compact) sides of a given 2-surface amounts to flipping the sign of the surface integral [$s_\mu \mapsto  - s_\mu$ in equation~\eqref{eq:MK}], and therefore necessarily leads to a zero total mass. However, using other definitions of mass, that property might not hold anymore. Of course, this argument is also valid in {\MyTheory}.

\subsection{Vacuum $\mS^3$ in General Relativity vs.  in {\MyTheory}}

In this section, we compare general relativity with {\MyTheory} in describing an empty $\mS^3$ universe while allowing for singularities, taking the example of the solution studied in Section~\ref{sec:The_solution}.

\subsubsection{With spherical symmetry}
\label{sec:Vac_GR_sphe}

{Let us consider the following setup:} a 3-sphere filled with a compact spherically symmetric patch of matter, and vacuum for the rest of the volume. We would expect this setup to be described by general relativity. Due to the Birkhoff theorem, if that setup is solution of the Einstein equation with $\Lambda=0$, then the vacuum region must be locally equivalent to the Schwarzschild metric, and more precisely, that whole region, by continuity, is described by a single Schwarzschild metric.\footnote{A similar situation, with two patches of matter instead of one, was studied by Sussman \citep{1985_Sussman,1986_Sussman}. In this study, the two patches of matter are homogeneous and correspond to FLRW spacetimes replacing the spatial infinities of the Schwarzschild metric.}
Placing the matter at the north pole, the south pole is by hypothesis a regular vacuum center of symmetry, around which the volume is finite. Such a center of symmetry is only possible with the Schwarzschild metric if a zero mass parameter is assumed. In other words, the metric in the vacuum region of a 3-sphere with one compact spherically symmetric distribution of matter is the Minkowski metric due to Birkhoff's theorem. {Still within GR,} this result holds regardless of the size of the patch of matter. Therefore, for a patch of matter small compared to the size of the 3-sphere, that sphere would be exactly spatially flat nearly everywhere except from this small patch. This is not a behaviour we would deem ``natural''. What would be expected for such a setup is to have a vacuum region having an isotropic positive spatial curvature, constant far from the patch of matter.\footnote{In reality, what probably happens for this setup is that the total volume of the flat region (i.e. the vacuum) should be small compared to the volume of the patch of matter, since all the curvature needed to retrieve the topology of $\mS^3$ is localised in the matter.} 
The situation is even ``worse'' if, instead of a patch of matter, we consider vacuum everywhere on the sphere except for eventual singularities (still assuming spherical symmetry). The Birkhoff theorem still applies, meaning that everywhere on the sphere but on one or two poles, the metric is locally equivalent to a single Minkowski metric (for one singularity), or to a single Schwarzschild metric (for two singularities). This is not possible if we require finite spatial volume induced by the metric \citep{2011_Uzan_et_al}. 

{The above discussion} is a quick explanation for why a spherically symmetric vacuum (but allowing for singularities) solution on $\mS^3$ of general relativity is not possible \citep[see][for a detailed explanation]{2011_Uzan_et_al}. In other words, general relativity cannot describe one BH (or two if placed at opposite poles) or repulsive naked singularities (the result does not depend on the sign of the Schwarzschild mass parameter) in an empty $\mS^3$ universe with $\Lambda=0$. As is well known, this also holds for a homogeneous vacuum in $\mS^3$ with or without~$\Lambda$.\saut

To our opinion, this result suggests the same interpretation we made from the result derived in \citep{2023_Vigneron_et_al_b}: i.e. that general relativity might be incomplete for non-Euclidean topologies. In \citep{2023_Vigneron_et_al_b}, that interpretation came from the result of the non-existence of a non-relativistic limit of general relativity in $\mR\times\mS^3$. In the present discussion, if we consider that one or two black holes in $\mS^3$ should be something authorised by our relativistic gravitation theory (regardless of the value of $\Lambda$), this means that we should modify the Einstein equation such that this physical system is authorised. It turns out that the modification proposed in~\citep{2023_Vigneron_et_al_b} to solve the ``problem'' related to the non existence of the non-relativistic limit, also solves the ``problem'' related to the non-existence of a vacuum, or a single/double singularity in $\mS^3$ in GR (for the regular vacuum case, it is given by our solution with $\chih = 0$). Of course the static singular solution discussed in the present paper features the unphysical repulsive singularity, but as a proof of concept, it shows that we can describe systems in non-trivial topologies that are not authorised by general relativity. The definite proof that a physical system with only attractive singularities in $\mS^3$ can be described by {\MyTheory} still remains to be given with an exact metric. This is discussed in Section~\ref{sec:BH_Alone}.

\subsubsection{Without spherical symmetry}
\label{sec:Lattice}

If we drop the hypothesis of one or two BHs, i.e. we drop spherical symmetry, it is possible to describe BHs in $\mS^3$ in GR using the ``lattice cosmology'' approach\footnote{It is possible to keep spherical symmetry and have a solution on $\mS^3$ if $\Lambda \not = 0$ is assumed \cite[see][]{2011_Uzan_et_al}, but two causally disconnected BHs are needed.}. This approach aims at describing BHs in an $\mS^3$ universe by solving exactly the Gauss--Codazzi constraints for the initial conditions \citep[e.g.][]{2012_Clifton_et_al, 2012_Bentivegna_et_al, 2014_Clifton, 2018_Bentivegna_et_al}, and then numerically evolving the solution. {This method is of importance for cosmology as it describes a cosmological model with a highly inhomogeneous distribution of matter, allowing for the study of the non-linear regime of structure formation \citep{2018_Bentivegna_et_al}. For this reason, a comparison between the implementation of this method in general relativity and in topo-GR is of interest. In the following discussion, we aim at presenting what difference we expect between lattice cosmology in GR and in topo-GR, taking into account the solution studied in this paper.}\saut

The principle of the method of lattice cosmology is to assume vacuum; initial staticity of the metric, i.e. $K_{ij} = 0$ for the initial extrinsic curvature\footnote{It is also possible to have initial non-staticity by assuming, e.g., $K_{ij} = \sqrt{\Lambda/3} \, h_{ij}$ \citep{2017_Durk_et_al_b}.}; and perform a conformal decomposition of the initial spatial metric assuming the conformal metric to be $\tilde h^{\mS^3}_{ij}$. In that case, the initial conformal factor $\phi$ is solution of a linear equation: the Lichnerowicz--York equation
\begin{align}
	\left(\Delta_{\mS^3} - \frac{\tilde\CR}{8}\right)\phi = 0, \label{eq:LichYork}
\end{align}
where $\tilde\CR$ is the scalar curvature of the conformal spatial metric. The BH setup on $\mS^3$, or ``punctures'' in the vocabulary used by \citep{2012_Bentivegna_et_al, 2014_Korzynski}, is then obtained by summing the solution of that equation with a single Dirac delta as a source, and placed at different positions on the sphere. The initial condition is then evolved numerically.\saut

The initial hypersurface described by the metric $h_{ij} = \phi^4\tilde h^{\mS^3}_{ij}$ is fully spacelike and is similar to the fully spacelike hypersurfaces we obtain with the Schwarzschild metric and the {BH-RNS} solution in isotropic coordinates, as described in Appendix~\ref{sec:Iso_Coord}. Similarly to Schwarzschild, that hypersurface in the case of lattice cosmology has infinite volume, and features horizons which delimit different causally disconnected regions. While the full volume is infinite, it can be shown that for~$N\geqslant3$ black holes, one of the regions has a finite volume and is surrounded by $N$ horizons (see for instance Figure~2 in \citep{2014_Clifton}). This region is interpreted as describing $N$ black holes in an expanding 3-sphere. The other regions are asymptotically flat (called ``infinite flat ends''), as is the case with the Schwarzschild metric. These regions represent an asymptotically flat space filled with one non spherically symmetric BH.\saut

{The {BH-RNS} solution of topo-GR has a finite spatial volume (see Figure~\ref{fig_Lattice_double}) as the infinities of the Schwarzschild metric have been replaced by repulsive naked singularities at finite distance. This suggests that the same behaviour will be present in a lattice cosmology approach of topo-GR: with any number of BHs, all regions should have a finite volume and could each be interpreted as an $\mS^3$ universe. In other words, we expect that the infinite flat ends present in lattice cosmology in GR will be replaced by repulsive naked singularities at finite distance in topo-GR.}  
{It remains to be shown that the lattice cosmology approach can be performed in topo-GR.} In particular, a linear equation similar to~\eqref{eq:LichYork} must be obtained. Such a study is an interesting follow-up to the work of the present paper.

\subsection{{Towards a non-static physical solution}}
\label{sec:BH_Alone}

{In this section, we discuss what is the next step in the search for exact inhomogeneous solutions of topo-GR. We also put in perspective this search with the study of BHs in an expanding closed universe.}

\subsubsection{Removing the staticity hypothesis}
\label{sec_non_stat}

{As presented in Section~\ref{sec:Why}, the initial hypothesis of staticity is probably what led to the presence of (non-physical) repulsive singularities. Therefore, a more physical approach would be to solve the system~\eqref{eq:biCoEq}--\eqref{eq:biCoCond} with the line element~\eqref{eq:non-static}, i.e. with time dependence. We expect the solution to represent one, or two, BHs (with positive mass) at opposite poles of an {\it expanding} 3-sphere.} As a first step towards finding this metric, Newtonian gravitation on $\mS^3$ can help us guessing its first order. The Newtonian gravitational potential of a point mass in an expanding vacuum $\mS^3$ universe was derived in \citep{2023_Vigneron_et_al_a} to be
\begin{align}
	\Phi(\chi, t) = -\frac{M_{\rm N}}{a(t)}\left(\cot\chi\right) \left(1-\chi/\pi\right), \label{eq:Phi_N}
\end{align}
where $a(t) = \left(\frac{3\kappa M_{\rm N}}{8\pi^2}\right)^{1/3} \, t^{2/3}$. Topo-GR is built such that it includes Newtonian gravitation in any spatial topology in a limit $c\rightarrow \infty$. Therefore, using the dictionary derived in \citep[][Section~6.3]{2023_Vigneron_et_al_b}, we can predict a possible solution of equations~\eqref{eq:biCoEq}--\eqref{eq:biCoCond} with the non static spherical line element~\eqref{eq:non-static} to be
\begin{align}
	\dd s^2 &= -\left[1 + 2\frac{M_{\rm N}}{a(t)\,  c^2}\left(\cot\chi\right) \left(1-\chi/\pi\right) \right]\dd t^2 \\
		&\qquad + a^2(t)\left\{\left[1 - 2\frac{M_{\rm N}}{a(t)\, c^2}\left(\cot\chi\right)  \left(1-\chi/\pi\right) \right]\dd\chi^2 + \sin^2\chi \, \dd\Omega^2\right\} + \bigO{1/c^3}, \nonumber
\end{align}
at leading orders. The full exact metric is expected to be only a function of the Newtonian potential~\eqref{eq:Phi_N} and the scale factor, as is often the case in solutions of general relativity built from Newtonian potentials.\saut

\subsubsection{A black hole in an expanding, finite volume, universe}

{The aforementioned metric would describe a black hole in an expanding Universe. The study of such a system is of high interest with respect to a recent debate on a possible coupling between black holes and the expansion of the Universe. This coupling was hypothesised by \citet{2019_Croker_et_al}, claimed to be observed by \citet{2023_Farrah_et_al_b}, and questioned by (e.g.) \citet[][]{2023_Cadoni_et_al}. Therefore, the interest of searching for such a metric goes beyond the study of topo-GR, as it will also enter the debate about the role of black holes in cosmology.}\saut

Most of the existing approaches for studying a black hole in an expanding universe assume: i) non-vacuum; ii) asymptotically infinite FLRW spacetime. To our opinion, a key complementary approach would be to rather consider the following set of hypothesis:
\begin{enumerate}[label=(\roman*)]
	\item vacuum,
	\item finite volume, non-static spacelike hypersurfaces.
\end{enumerate}
The vacuum hypothesis may be more representative of our Universe which can be seen as being filled with small patches of matter surrounded by vacuum. Because of the vacuum hypothesis, the evolution of the matter density at infinity cannot be used to track the expansion anymore. In other words, without matter, there is no reference allowing for the definition of expansion. In a sense, this is in line with the interpretation of Mach's principle made by Einstein. Therefore, the role of the volume finiteness hypothesis is to reintroduce such a reference and to allow for the study of the expansion via the evolution of the global compact spacelike volume surrounding the BH. {The non-static metric we {would look for}, as described in Section~\ref{sec_non_stat}, {would fulfill} these two hypotheses. In this sense, this metric would offer a complementary approach in studying the relation between black holes and expansion.}

\section{Conclusion}
\label{sec:Conclusion}

In this paper, we derived a first exact inhomogeneous solution in a non-Euclidean topology (i.e. for which the spatial covering space in not $\mE^3$) of the Topologically Corrected General Relativity ({\MyTheory}) developed in \citep{2024_Vigneron}. The solution is static with spherical symmetry and represents an attractive singularity shielded by a horizon (black hole) and a repulsive naked singularity placed at opposite poles of a 3-sphere. We analysed the properties of this solution. In particular, we showed that its Penrose--Carter diagram is {similar to} the Schwarzschild diagram, but from which the two spatial infinities are cut and replaced by {repulsive naked singularities} at finite distance, and that the total Komar mass of the solution is zero: positive mass for the black hole and opposite negative mass for the {repulsive naked singularity}. This zero total mass is consistent with an important result of {\MyTheory} derived in \citep{2023_Vigneron_et_al_b} for cosmology: a closed universe filled with homogeneous matter cannot be static unless the total mass is zero. While that result was derived for a homogeneous solution, the metric obtained in the present paper confirms this result for a specific inhomogeneous solution. {This suggests that any physical solutions (i.e. with positive mass) of topo-GR in closed topologies require non-staticity.}\saut

{A direct follow-up we expect for this paper is to search for a non-static solution that would describe a single black hole in an expanding universe with the topology of a 3-sphere, something which is not possible in general relativity. To further identify the differences between topo-GR and general relativity in closed topologies, {and to obtain a highly inhomogeneous model universe for cosmology in this context}, we also argued that an interesting study would be to develop the method of lattice cosmology for topo-GR.}

\section*{Acknowledgements}

%\begin{multicols}{2}
{\small
\noindent We thank Miko\l aj Korzy\'nski, Julien Larena, Fernando Piza\~na, Boudewijn Roukema and Roberto Sussman for insightful discussions. 

\noindent Q.V. and \'A.S. are supported by the Centre of Excellence in Astrophysics and Astrochemistry of Nicolaus Copernicus University in Toru\'n. 
\noindent Q.V. is supported by the Polish National Science Centre under Grant No. SONATINA 2022/44/C/ST9/00078. 
\noindent \'A.S. is supported in part by the Polish National Science Centre under Grant No. OPUS 2021/41/B/ST9/00757. 
\noindent P.M. is supported by the Spanish Ministerio de Ciencia e Innovaci\'on (Fondos MRR) --- Conselleria de Fons Europeus, Universitat i Cultura with funds from the European Union NextGenerationEU (PRTR-C17.I1) through the project ‘Tecnolog\'ias avanzadas para la exploraci\'on del universo y sus componentes’ (ref. SINCO2022/6719).
P.M.'s work was also supported by the Universitat de les Illes Balears (UIB); the Spanish Agencia Estatal de Investigaci\'on grants PID2022-138626NB-I00, RED2022-134204-E, RED2022-134411-T, funded by MICIU/AEI/10.13039/501100011033 and the ERDF/EU; and the Comunitat Aut\`onoma de les Illes Balears through the Servei de Recerca i Desenvolupament and the Conselleria d'Educaci\'o i Universitats with funds from the Tourist Stay Tax Law (PDR2020/11 - ITS2017-006), and from the European Union - European Regional Development Fund (ERDF) (SINCO2022/18146).
}

%\end{multicols}

%\newpage
\appendix

\section{Detailed derivation of the solution}
\label{app:Details}

In this section, all the parameters $c_i$ will be constants, and for simplicity we denote the variable $\tilde\chi$ related to the coordinates~\eqref{eq:pingouin} by $\chi$. \saut

We start with the ansatz
\begin{align}
	\dd s^2 = -A(\chi)\dd t^2 + \frac{c_0}{A(\chi)}\dd\chi^2 + C(\chi) \sin^2(\chi)\,\dd\Omega^2. \label{eq:ansatz}
\end{align}
Since we consider vacuum, therefore, solving equation~\eqref{eq:biCoEq} automatically solves the bi-connection equation~\eqref{eq:biCoCond}. The $00$-component of~\eqref{eq:biCoEq} with ansatz~\eqref{eq:ansatz} leads to
\begin{align}
	\frac{A(\chi)A'(\chi)}{2c_0} \left[\frac{C'(\chi)}{C(\chi)} + 2\cot\chi  + \frac{A''(\chi)}{A'(\chi)}\right] = 0,\nonumber
\end{align}
whence
\begin{align}
	C(\chi) = \frac{c_1}{A'(\chi) \, \sin^2\chi}
\end{align}
for a non trivial solution. Then, reinjecting this solution in ansatz~\eqref{eq:ansatz}, the $11$-component of equation~\eqref{eq:biCoEq} becomes
\begin{align}
	\frac{A'''(\chi)}{A'(\chi)} - \frac{3}{2}\left(\frac{A''(\chi)}{A'(\chi)}\right)^2 - 2 = 0,\nonumber
\end{align}
whose solution is
\begin{align}
	A(\chi) = - c_3 \, \cot\left(\chi + c_2\right) + c_4.
\end{align}
Then, reinjecting again this solution in ansatz~\eqref{eq:ansatz}, the $22$-component of equation~\eqref{eq:biCoEq} becomes
\begin{align}
	\cos\left(2\chi\right)\left[1 -\frac{c_1 c_4}{c_0 c_3}\cos\left(2c_2\right) - \frac{c_1}{c_0}\sin\left(2c_2\right)\right] + 
	\sin\left(2\chi\right)\left[ \frac{c_1 c_4}{c_0 c_3}\sin\left(2c_2\right) - \frac{c_1}{c_0}\cos\left(2c_2\right)\right] = 0. \nonumber
\end{align}
Since this equation must hold for any $\chi$, the two main terms of this equation are independently zero, which leads to
\begin{align}
	c_4 = c_3 \cot\left(2c_2\right) \quad ; \quad c_1 = c_0 \sin\left(2c_2\right).
\end{align}
 The $33$-component of equation~\eqref{eq:biCoEq} is equivalent to the $22$-component, and, therefore, is fulfilled. We are left with the solution
\begin{align}
	\dd s^2	&= - c_3\left[\cot\left(2c_2\right) - \cot\left(c_2+\chi\right)\right]\dd t^2 + \frac{c_0/c_3}{\left[\cot\left(2c_2\right) - \cot\left(c_2+\chi\right)\right]}\dd\chi^2 \nonumber\\
			&\qquad + \frac{c_0}{c_3}\sin\left(2c_2\right)\, \sin^2\left(\chi + c_2\right)\dd \Omega^2.\nonumber
\end{align}
This line element is not defined for $c_2 = 0 \ ({\rm mod} \, \frac{\pi}{2})$, but can be extended to those values by the change $c_5 \coloneqq c_3/\sin\left(2c_2\right)$, leading to
\begin{align}
	\dd s^2	&= - c_5\frac{\sin(\chi - c_2)}{\sin (\chi + c_2)}\dd t^2 + \frac{c_0}{c_5}\frac{\sin(\chi + c_2)}{\sin (\chi - c_2)}\dd\chi^2 + \frac{c_0}{c_5}\, \sin^2\left(\chi + c_2\right)\dd \Omega^2.\nonumber
\end{align}
This line element is $\pi$-periodic in the $c_2$-parameter and is invariant by the change $\left(\chi, c_2\right) \rightarrow  (\pi - \chi; \pi - c_2)$. This change corresponds to positioning the black hole either at the north or the south pole of the 3-sphere. Therefore, this solution is equivalent for $c_2 \in [0, \pi/2]$ and $c_2 \in [\pi/2, \pi]$. Finally, because the $\theta$ and $\varphi$ directions are spacelike, then, in our convention of signature $c_0/c_5 > 0$. We introduce $\Rc^2 \coloneqq c_0/c_5$ and $\chih \coloneqq 2 c_2$, which leads to the line element
\begin{align}
	\dd s^2	&= - c_5\frac{\sin(\chi - \chih/2)}{\sin (\chi+\chih/2)}\dd t^2 +  \frac{\sin(\chi+\chih/2)}{\sin(\chi - \chih/2)} \Rc^2\dd\chi^2 + \Rc^2 \sin^2(\chi +\chih/2)\,\dd \Omega^2, \label{eq:tagada}
\end{align}
where $\chih \in [0,\pi]$. The $c_5$-prefactor corresponds to the normalisation of time (constant lapse) and can be dropped. We obtained the line element~\eqref{eq:the_solution_har}, where we used the notation $\tilde\chi$ instead of $\chi$. We recall that this line element is solution of equation~\eqref{eq:biCoEq} in coordinates where the reference curvature has the form~\eqref{eq:Rbar_munu_init}.

\section{Compact spatial isotropic coordinates}
\label{sec:Iso_Coord}

%\begin{figure}[t]
%	\centering
%	\captionsetup{width=.9\linewidth}
%	\begin{subfigure}{0.49\textwidth}
%		\includegraphics[width=7cm]{Lattice_M=025_Sch.pdf}
%	\end{subfigure}
%	\begin{subfigure}{0.49\textwidth}
%		\includegraphics[width=7cm]{Lattice_M=025.pdf}
%	\end{subfigure}
%	\caption{2-dimensional section of the fully static spacelike hypersurface, in the case of the Schwarzschild metric (left), and our solution with $\chih = \pi/4<\pi/2$ (right). The horizon is represented as a black thick line. In both cases, region I' is on the left of the horizon, and region I on the right.\label{fig:3D}}
%\end{figure}

\begin{figure}[t]
	\centering
	\captionsetup{width=.9\linewidth}
	\begin{subfigure}{0.49\textwidth}
		\includegraphics[width=7cm]{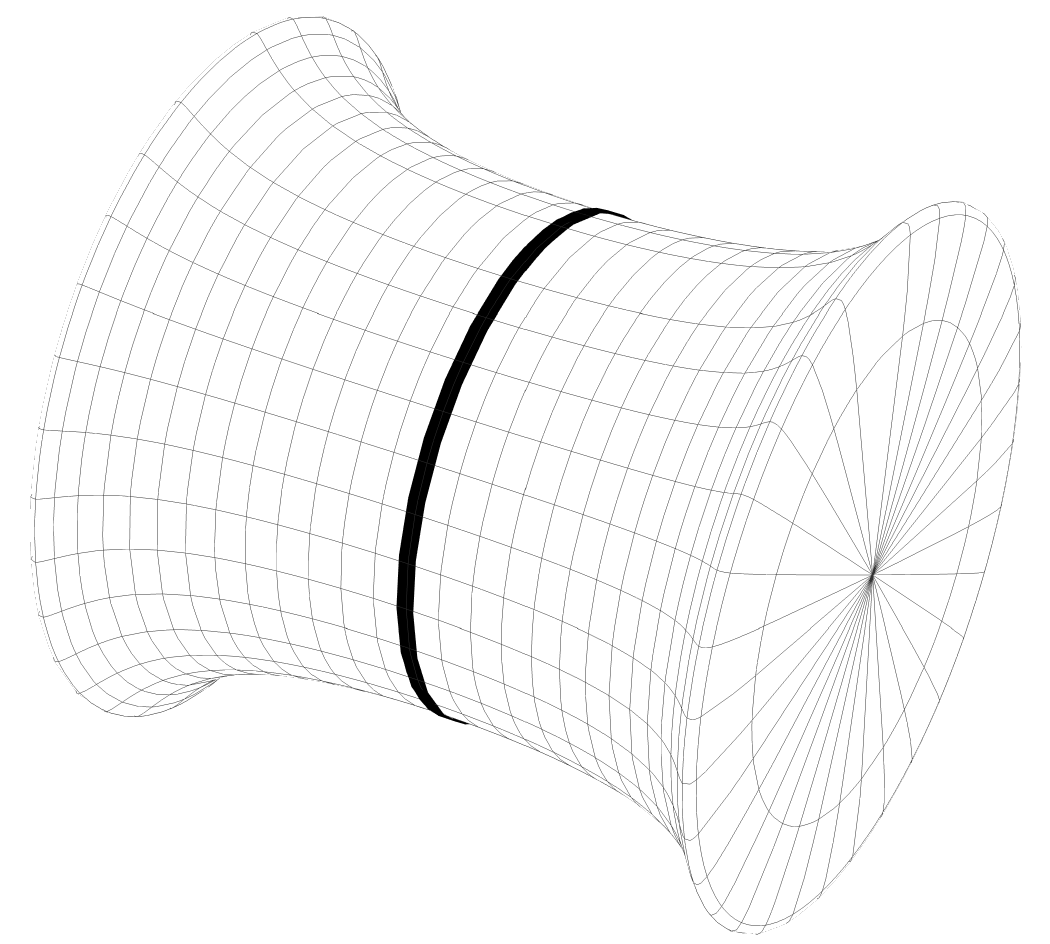}
	\end{subfigure}
	\begin{subfigure}{0.49\textwidth}
		\includegraphics[width=7cm]{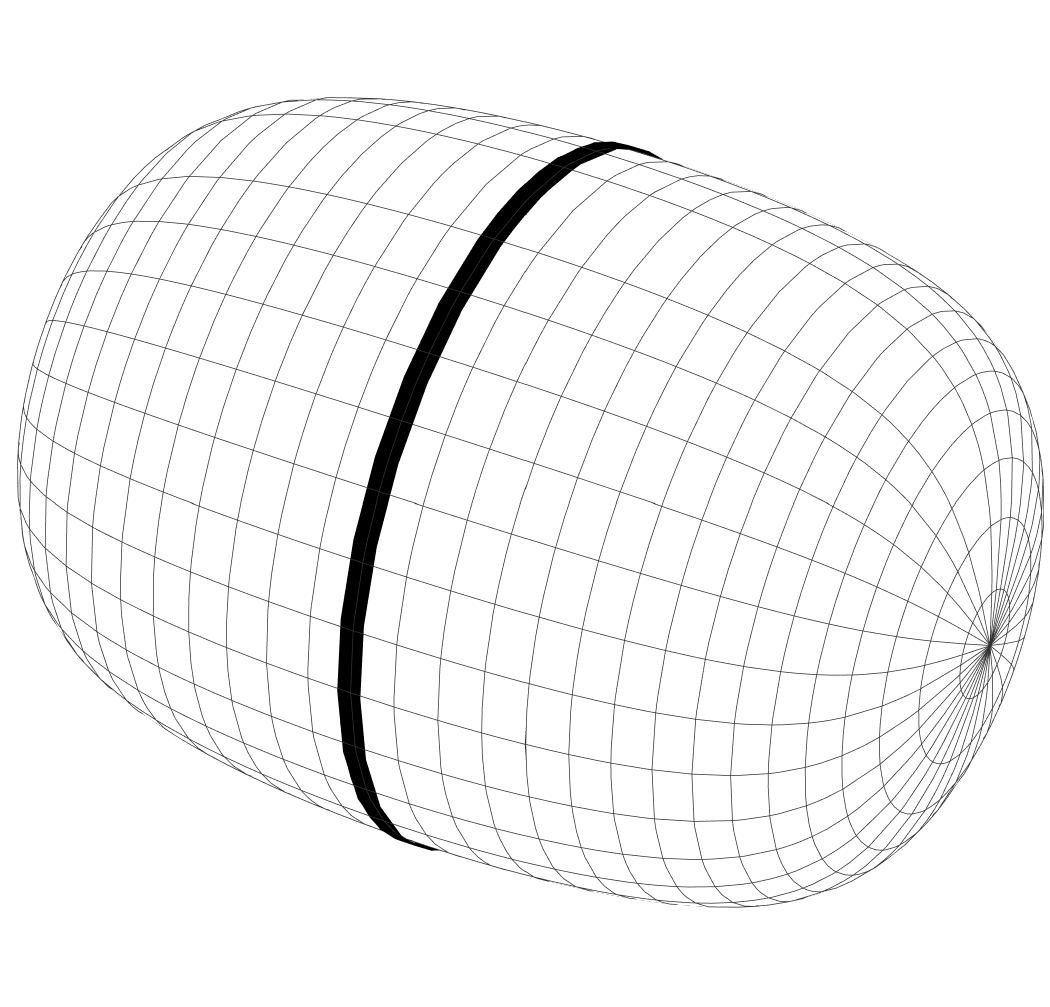}
	\end{subfigure}
	\caption{2-dimensional section of the fully static spacelike hypersurface of the {BH-RNS} solution with $\chih = \pi/4<\pi/2$ (left) and $\chih =  3\pi/4>\pi/2$ (right). The horizon is represented as a black thick line. In both cases, region I' is on the left of the horizon, and region I on the right.\label{fig:3D}}
\end{figure}

%\begin{figure}[t]
%	\centering
%	\captionsetup{width=.9\linewidth}
%	\includegraphics[width=7cm]{Lattice_M=075.pdf}
%	\caption{2-dimensional section of the fully static spacelike hypersurface of our solution for $\chih =  3\pi/4>\pi/2$.\label{fig:3Dbis}}
%\end{figure}

We derive in this section the compact spatial isotropic coordinates related to the metric~\eqref{eq:the_solution}. The hypersurfaces of the foliation adapted to these coordinates are spacelike everywhere and of finite volume.\saut

For the Schwarzschild metric, the standard coordinate system describes the static region I and the non-static region II, without including regions I' and II'. There exists a coordinate system encompassing smoothly the two static regions I and I', but without II and II'. This is the so-called isotropic coordinate system, which, when written with a compactification of the radial coordinates, leads to the following Schwarzschild line element
\begin{align}
	\dd s^2_{\rm Sc} = -\frac{1-\tan\frac{\tilde\chi}{2}}{1+\tan\frac{\tilde\chi}{2}}\dd t^2 + \frac{R_{\rm S}^2}{64}\left(\frac{1}{\cos\frac{\tilde\chi}{2}} + \frac{1}{\sin\frac{\tilde\chi}{2}} \right)^4\left[\dd\tilde\chi^2 + \sin^2\tilde\chi\,\dd\Omega^2\right], \label{eq:SC_Iso}
\end{align}
where $R_{\rm S}$ is the Schwarzschild radius and $\tilde\chi \in ]0,\pi[$. A representation in this coordinate system of the $\theta = \pi/2$ section of the corresponding (spacelike) $t={\rm cst}$ hypersurfaces is presented as translucent white in Figure~\ref{fig_Lattice_double}. The horizon is represented as a black thick line and separates regions I' (on the left of the horizon) to I (on the right of the horizon).\saut

The equivalent coordinate system in the case of the solution~\eqref{eq:the_solution} is obtained with a coordinate transformation $\chi \rightarrow \tilde\chi(\chi)$ such that the line element has the form
\begin{align}
	\dd s^2 = -\frac{\sin\left[\chi(\tilde \chi) - \chih\right]}{\sin\left[\chi(\tilde \chi) \right]} \dd t^2 + f(\tilde\chi)\left[\dd\tilde\chi^2 + \sin^2\tilde\chi\,\dd\Omega^2\right], \label{eq:iso_BHWH}
\end{align}
where $f(\tilde\chi) =\sin^2\left[\chi(\tilde\chi)\right]/\sin^2\tilde\chi$ is a spatial conformal factor. In this coordinate system, the spatial metric is conformally equivalent to the homogeneous and isotropic metric on~$\mS^3$. The coordinate transformation required to obtain this line element is solution of
\begin{align}
	\chi'(\tilde\chi)^2 = \frac{\sin\left[\chi(\tilde \chi) - \chih\right]\sin\left[\chi(\tilde\chi)\right]}{\sin^2\tilde\chi}.
\end{align}
We assume $0<\tilde\chi<\pi$ (compactification of the coordinates), and require $\tilde\chi(\chi = \chih) = \pi/2$ [this places the horizon at $\tilde\chi = \pi/2$, as is the case with the Schwarzschild line element~\eqref{eq:SC_Iso}]. We obtain
\begin{align}
	\chi(\tilde\chi) = 
	\begin{dcases}
		F^{-1}\left[\ln\left(\tan\frac{\tilde\chi}{2}\right) | \chih\right] \qquad&{\rm for}\quad \frac{\pi}{2} < \tilde\chi <2 \, \arctan\left[ e^{F(\pi | \chih)}\right] \\
		F^{-1}\left[\ln\left(\cot\frac{\tilde\chi}{2}\right) | \chih\right] \qquad&{\rm for}\quad 2 \, {\rm arccot} \left[ e^{F(\pi | \chih)}\right] < \tilde\chi < \frac{\pi}{2}
	\end{dcases},
\end{align}
where
\begin{align}
	F\left(\chi|\chih\right) \coloneqq \int_0^{\chi-\chih} \frac{1}{\sqrt{ \sin\left(x+\chih\right)\sin x}} \dd x,
\end{align}
and $F^{-1}\left(\chi|\chih\right)$ is the inverse function of $F\left(\chi|\chih\right)$ with respect to the first variable.\saut

A section $\theta = \pi/2$ of the spatial hypersurfaces described by this coordinate system is presented on the left panel of Figure~\ref{fig:3D}, for $\chih = \pi/4$. We see that this section is similar to the one of the Schwarzschild metric, but instead of having an asymptotically infinite flat hypersurface for $\tilde\chi = 0$ and $\tilde\chi = \pi$, the section is closed at those two positions (on which the two {RNSs} are present). This illustrates the closure of the solution's Penrose--Carter diagram diagram at the former spatial infinities, with respect to the Schwarzschild case.\saut

In the case $\chih > \pi/2$, the surface of the $\mS^2$-sections encompassing the horizon is strictly decreasing from the horizon to the {RNSs}, as can be seen in the right panel of Figure~\ref{fig:3D} (corresponding to $\chih = 3 \pi/4$).

\section{Komar mass in {\MyTheory}}
\label{app_Komar}

We show in this section that, in {\MyTheory},  the Komar mass, as defined by equation~\eqref{eq:MK}, is still invariant up to changes of the 2-surface $\CS_t$ in a vacuum region (i.e. $T_{\mu\nu} = 0$), provided either the Killing vector $k^\mu$ or the normal vector $n^\mu$ to $\Sigma_t$ are in the kernel of $\bar R_{\mu\nu}$. Those conditions are fulfilled by the {BH-RNS} metric.\saut

We consider a compact volume $\CV_t$ in $\Sigma_t$ not containing any singularity and delimited by two 2-surfaces $\CS_t$ and $\CH_t$. Following the derivation made in Section~8.6.1 in \citep{2012_GG}, the Komar mass $M_K^{\CS_t}$ related to $\CS_t$ is
\begin{align}
	M_K^{\CS_t}
		&\coloneqq -\frac{1}{\kappa}\int_{\CS_t} \left(s_\mu n_\nu - n_\mu s_\nu\right)\nabla^\mu k ^\nu\sqrt{q} \, \dd^2 y \nonumber \\
		&=  \frac{2}{\kappa}\int_{\CV_t} R_{\mu\nu} k^{\mu} n^\nu \sqrt{{\rm det}\, h_{ij}} \, \dd^3 x + M_K^{\CH_t}, 
\end{align}
where $M_K^{\CH_t}$ is the Komar mass related to $\CH_t$. Then, using the field equation~\eqref{eq:biCoEq} of {\MyTheory}, we obtain
\begin{align}
	M_K^{\CS_t}
		&=  \frac{2}{\kappa}\int_{\CV_t} \left[\bar R_{\mu\nu}  + \kappa \left(T_{\alpha\beta} - \frac{T}{2} g_{\alpha\beta} \right)\right]k^\mu n^\nu \sqrt{{\rm det}\, h_{ij}} \, \dd^3 x + M_K^{\CH_t}, 
\end{align}
putting $\Lambda$ in $T_{\mu\nu}$. Therefore, if there is vacuum on $\CV_t$, i.e. $T_{\mu\nu}=0$ in $\CV_t$, and $\bar R_{\mu\nu} k^\mu=0$ or $\bar R_{\mu\nu} n^\nu = 0$, which is the case for the calculation made in Section~\ref{sec:Mass}, then $M_K^{\CS_t} = M_K^{\CH_t}$. The Komar mass in {\MyTheory} is thus, under the conditions stated above, independent on the choice of the 2-surface in vacuum. 
\hfill$\square$

\bibliographystyle{QV_mnras}
\bibliography{bib_BH-NWH}

\end{document}